\documentstyle[12pt,aaspp4]{article}
 
\lefthead{Spitzak and Schneider}
\righthead{Properties of HI--Selected Galaxies}
 
\begin{document}
 
\title{ A Deep Survey of HI--Selected Galaxies:\\ The Sample and the Data }

\author{John G. Spitzak\altaffilmark{1,2} and Stephen E. Schneider\altaffilmark{1}}
\affil{Astronomy Program, University of Massachusetts,
    Amherst, MA 01003}
 
\altaffiltext{1}{Visiting Astronomer, Kitt Peak National Observatory,
National Optical Astronomy Observatories (NOAO), which is operated by the
Association of Universities for Research in Astronomy, Inc. (AURA)
under cooperative agreement with the National Science Foundation.}
\altaffiltext{2}{present address: Sea Beam Instruments, Boston, MA}
 
\begin{abstract}

In a 21 cm neutral hydrogen survey of approximately 55 deg$^2$ out to a
redshift of $cz=8340$ km s$^{-1}$, we have identified 75 extragalactic HI
sources.
These objects comprise a well-defined sample of extragalactic sources
chosen by means that are independent of 
optical surface brightness selection effects.
In this paper we describe the Arecibo survey procedures and HI data,
follow-up VLA HI observations made of several unusual sources, and Kitt
Peak $B$-, $R$-, and $I$-band photometry for nearly all of the galaxies.
We have also gathered information for some of the optically detected
galaxies within the same search volume.

We examine how samples generated by different types of search techniques
overlap with selection by HI flux.
Only the least massive HI object, which is among the lowest mass HI sources
previously found, does not have a clear optical counterpart,
but a nearby bright star may hide low surface brightness emission.
However the newly-detected systems do have unusual optical properties.
Most of the 40 galaxies that were not previously identified in
magnitude-limited catalogs appear to be gas-dominated systems,
and several of these systems have HI mass-to-light ratios among the
largest values ever previously found.
These gas-dominated objects also tend to have very blue colors, low surface
brightnesses, and no central bulges, which correlate strongly with
their relative star-to-gas content.

\end{abstract}
 
 
\section{Introduction}
 
A strong argument can be made that the present census of galaxies is
biased toward objects which are intrinsically bright and easy to
identify optically (Disney \& Phillipps 1983; McGaugh 1996). Extragalactic
astronomy's historical dependence on searches of optical plates for galaxy
identification almost certainly has led to an under-count of low surface
brightness (LSB) objects. This bias leads to an
under-representation of these objects and their properties when examining
the overall extragalactic population---information which is essential when
investigating the evolution of galaxies or the large-scale structure and
fate of the Universe. Ridding the extragalactic census of its optical bias,
or even assessing its seriousness is problematic.  In this paper we present
a sample of HI-selected extragalactic sources that permits us to
examine the gaseous and stellar properties of galaxies in a fresh light.

Most measurable galaxy properties are in some way a direct result of
stellar emission, stellar remnants, or the processes that form stars. 
For example, far-infrared emission from galaxies is essentially starlight
re-radiated by dust grains, and the dust grains' existence is a result of
stellar evolution.  Far infrared emission is therefore highly correlated
with the optical emission of stars, and it is of limited utility in
identifying objects that cannot easily be detected optically.

One strategy that has been used to correct the optical bias is to search for
extragalactic sources at faint surface brightnesses. This was effectively
begun by Nilson (1973) in the Uppsala General Catalog (UGC), where the 
criterion for acceptance was a minimum angular size instead of a magnitude
limit. Since galaxies' intrinsic diameters vary by a much smaller factor
than do their luminosities, a diameter-limited survey detects a larger
proportion of LSB galaxies. The UGC is, of course, also limited
by the surface brightness sensitivity of the Palomar Observatory Sky
Survey (POSS) plates (about 25 mag arcsec$^{-2}$). This has been pushed to
fainter surface brightnesses by using the second generation (POSS II)
plates (Schombert \& Bothun 1988; Schombert et al.~1992; Pildis, Schombert,
\& Eder 1997). Special photographic amplification techniques that go even
deeper have revealed some remarkable LSB sources like Malin 1
(Bothun et al.~1987), which has one of the largest known HI masses.
However, seeking sources at ever fainter
surface brightnesses becomes more and more difficult and it ultimately
remains tied to the presence of starlight. It would be preferable
to take an independent approach to sampling the extragalactic population.

A galaxy's atomic hydrogen (HI) gas content is one of the few properties
that should be relatively uncorrelated with its optical emission.
Of course, like any extensible property of a galaxy, some degree of
correlation is expected due to the galaxy's total bulk. However, since
hydrogen is primordial, it does not require
star formation for its existence, and since the 21 cm line's excitation
temperature is so low, starlight is not needed to excite it. Its status as the
``raw material'' in star formation makes it likely that HI gas content
and stellar emission will be linked in many galaxies, but since the gas is
consumed in the process, the amounts may be anticorrelated. For example,
it is well known that elliptical galaxies may contain
little or no measurable HI, and we might also expect to find reservoirs of
HI where star formation has been inefficient so that the HI emission is
more readily detectable than the optical emission.

Some intriguing sources with no optical counterpart have been discovered
by serendipity at 21 cm, like the ring of HI gas in Leo (Schneider 1989)
and the southwest clump of the Virgo cluster cloud HI 1225+01 (Giovanelli,
Williams, \& Haynes 1991). The northeast clump of HI 1225+01 has a
ratio of $M_{HI}/L_B=10$ (Salzer et al.~1991), and some other extreme
objects, like DD0 154 and Malin 1, have $M_{HI}/L_B\approx5$ (Carignan \&
Beaulieu 1989; Bothun et al.~1987).
These objects represent the gas-rich extreme of a continuum of extragalactic
sources that also ranges to objects that are composed almost
entirely of stars, like ellipticals and dwarf spheroidal systems.
The enormous range of $M_{HI}/L$ found in
extragalactic sources indicates that HI properties of extragalactic
sources cannot be fully appreciated from optically-selected samples.

To generate an HI-selected sample of galaxies, we have conducted a
large, sensitive survey for the extragalactic 21 cm emission using the
Arecibo radio telescope.\footnote{The Arecibo Observatory is part
of the National Astronomy and Ionosphere Center, which is operated by
Cornell University under cooperative agreement with the National Science
Foundation. in Puerto Rico.}
Our ``Arecibo Slice'' covers about 1$^\circ$ in
declination and a total area of about 55 sq deg out to a redshift
of $cz_{hel}=8340$ km s$^{-1}$.
We also examine the literature for optically-selected sources that
reside in our survey volume. Most of these were ``rediscovered''
by the 21 cm survey, but we have attempted to track down other galaxies
from magnitude-limited and other types of catalogs.

In addition, we have collected multi-wavelength optical data to allow
color and luminosity comparisons.
We have also observed several of the most unusual galaxies detected
in the survey with the VLA\footnote{The National Radio Astronomy Observatory
is a facility of the National Science Foundation operated under cooperative
agreement by Associated Universities, Inc.} in order to examine their
HI sizes, and better-determine their positions.
In this paper we present these
data and examine the differences, overlaps, and similarities of the
gaseous and stellar properties of our sample galaxies.

Our survey has been more successful than previous searches at identifying
low mass HI sources and has found a substantial population of objects
whose gas content dominates over their stellar content.
Among our 75 HI-selected sources is one of
the lowest mass field HI sources found to date, with
an HI mass similar to M81 dwarf A found by Lo \& Sargent (1979) 
in an HI survey of that nearby group.
Based on optical data collected at Kitt Peak, 
seven of our objects have $M_{HI}/L_B>3$, which was the value found
for M81 dwarf A (Sargent, Sancisi, \& Young 1983).  A dozen of
our objects also have substantially bluer colors than any of the
HI-detected galaxies studied by Szomoru et al.~(1994), or in
diameter-limited samples of galaxies like de Jong \& van der Kruit (1994).

\begin{figure}[tb]
\plotone{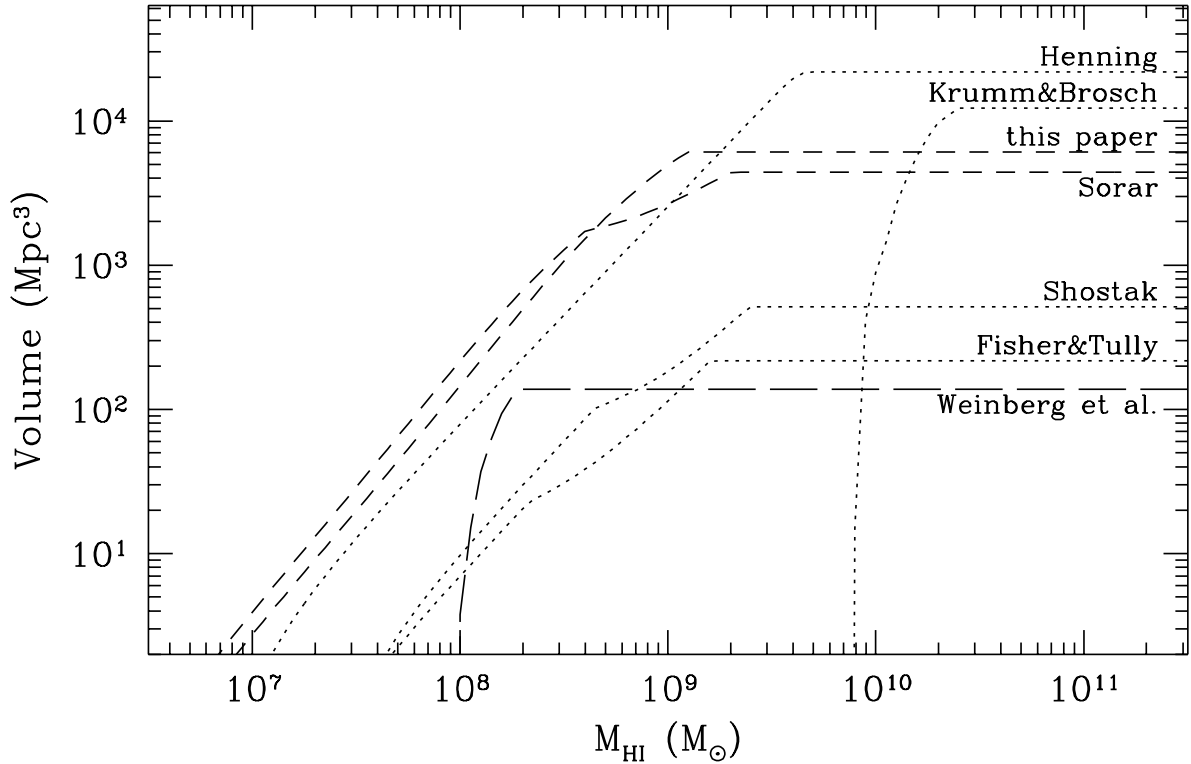}
\caption{Comparison of mass--volume sensitivity of various ``blind''
HI surveys.  The lines indicate the approximate volume within which 
a source of a given HI mass could have been detected based on the survey's
resolution, bandpass, noise level, and areal coverage. See Schneider
(1997) for details of the different surveys.
}
\label{surveys}
\end{figure}

In large part, this survey's success results from the size of the
volume examined.
This is a useful starting point for comparing surveys and is illustrated in
Fig.~\ref{surveys}. We show the volume within which various ``blind
surveys'' (not based on earlier optical imaging)
have been potentially sensitive to an HI source of a given mass.
This is an approximation based on the area examined, the quoted $rms$ noise,
redshift range covered, and sensitivity across the bandpass for each
survey (see Schneider 1997 for more details about the surveys).
This graph assumes the HI sources all have the same velocity width, and
that the search techniques are all equivalent. In fact, though, detection
rates are influenced by a wide variety of factors that can be
difficult to quantify.

Early surveys with the Green Bank 300 foot telescope by Shostak (1977)
and Fisher \& Tully (1981) were limited
by receiver sensitivity and the total bandpass available. 
Krumm \& Brosch (1984) surveyed a large volume by searching at
higher redshifts, but they were then insensitive to low mass sources.
With improved receiver systems, Henning (1990, 1995) was able to search
within larger volumes over a wide range of masses. 
The Green Bank 300 foot telescope was in many ways an ideal
survey instrument because the main limitation to search efficiency is,
in principle, the system temperature of the receiver system
(see, for example, Schneider 1996),
which was excellent at the 300 foot telescope.

A smaller-beam telescope has the advantage of avoiding confusion,
which is important because of galaxies' tendency to cluster.
This is particularly demonstrated by a survey conducted using the VLA
(Weinberg et al.~1991; Szomoru et al.~1994), which searched a volume
comparable to some of the early surveys. They identified nine previously
uncataloged sources, but six of these were in the neighborhood of bright
cataloged galaxies and might well have been assumed to be part of these
galaxies if surveyed with the $\sim11'$ beam of the Green Bank 300 foot
telescope.
The VLA also has advantages for interference rejection, but sensitivity
and bandpass limitations at the VLA constrain the amount of volume it can
search. The restriction on bandpass is the reason why the sensitivity of the
VLA survey drops off sharply below $10^8 M\sun$.

The Arecibo radio telescope represents a good compromise. Its $\sim3.3'$
beam would have confused only one of the VLA-identified sources,
and it provides a wide bandpass with adequate spectral resolution to
detect narrow-line HI sources.
As shown in Fig.~\ref{surveys} we examined a substantially larger
volume at low HI masses than any of the Green Bank or VLA surveys.
A recent ``Arecibo HI Strip Survey'' or AHISS by Sorar (1994; also
see Zwaan et al.~1997) used a drift scan strategy to sample a
similar-sized volume. Not counting sidelobes, that survey
actually covered about a third as much area on the sky,
but the quoted $rms$ noise levels suggest it should
be sensitive over a slightly larger volume for low mass HI sources.
The methodology of our survey is different from the AHISS, which might
account for our higher detection rate of low-mass HI sources
despite the nominally smaller search volume. This is a crucial point
for determination of the HI
mass function, and we discuss it at greater length in a subsequent
paper (Schneider, Spitzak, \& Rosenberg 1998) where we apply tests
to determine sample completeness.

In this paper we begin in \S 2 with a description of the survey
techniques and the Arecibo and VLA HI data. In \S 3 we describe the
optical data and cross-identifications of our sources with other catalogs.
In \S 4 we focus on distance-independent properties of the galaxies, with
a particular interest in understanding the selection effects that
determine which galaxies are included in which types of surveys.
Finally, in \S 5 we conclude with a summary of our results,
highlighting some of the most unusual individual objects that deserve
follow-up observations. We will analyze the intrinsic
distance-dependent properties of the galaxies, including their
luminosity function, in subsequent papers.

\section{The HI Survey}

\subsection{21 cm Observations}

The 21 cm survey observations were conducted at the Arecibo radio telescope
using the ``22 cm'' feed.  The Arecibo telescope has the largest collecting
area and highest resolution of any single-dish 21 cm telescope. 
The telescope response has a strong zenith-angle dependence, 
and our search strategy was largely dictated by this limitation. 
In addition, because this survey would require a large amount of observing
time with little idea of the detection rate, we selected a right ascension
range where there was relatively low ``proposal pressure'': 
the portion of the southern Galactic cap accessible from Arecibo.
Optically-identified sources in this region have been extensively studied
by Giovanelli \& Haynes (1993, and references therein), leaving
relatively less demand for telescope time from $22^h$ to $4^h$.

To optimize the survey we also needed to observe a region within the
10$^\circ$ zenith angle of the telescope's maximum sensitivity. And to
provide some transit time at full sensitivity this could be at most
about $8^\circ$. At the same time, we wanted to avoid
the $\sim3^\circ$ closest to zenith where the telescope can have difficulty
tracking a source. For the observatory's $18^\circ21'$ latitude, this gave
us a choice of $21^\circ21' < \delta < 26^\circ21'$ or
$10^\circ 21'<\delta<15^\circ21'$. Although the southern choice has the
advantage of being farther from the Galactic plane, we chose the northern
side because much more HI data had been published in that area
(Giovanelli \& Haynes 1989; Giovanelli et al. 1986). 
These previous HI observations allow us to test our completeness.

\begin{figure}[tb]\plotone{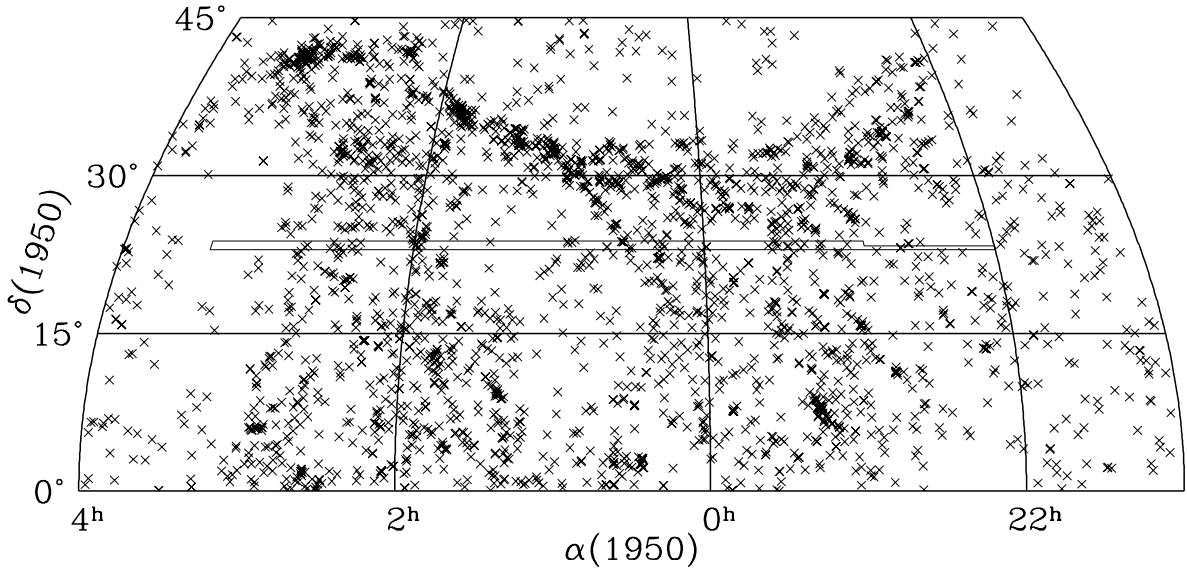}
\caption{Region surveyed in the Arecibo HI slice.
Positions of galaxies in the RC3 with redshifts in the range of
the survey (100--8340 km s$^{-1}$) are marked by $\times$ symbols.
The region around $\delta=23^\circ$ covered by the HI survey is shown
by an outline.
The serpentine ridge north of the slice is the Pisces--Perseus supercluster
at approximately 5000 km s$^{-1}$, and the higher density of galaxies
around $\alpha=2^h$ is part of the supergalactic plane.
\label{surveyarea}
}\end{figure}

The ``Arecibo Slice'' is within the region
$22^h 00^m < \alpha(1950)<  03^h 24^m$ and
$22^\circ 58' <\delta(1950)< 23^\circ 47'$, although in the portion
of right ascension earlier than $22^h 54'$, the survey reached a
maximum declination of $23^\circ19'$ (Fig.~\ref{surveyarea}).
Overall, close to $55$ deg$^2$ were surveyed in 14,130
pointings of the telescope. The positions observed were separated
by $4.1'$ on a hexagonal (``honeycomb'') grid. 
A hexagonal grid is preferable to a rectangular grid because there
is less sensitivity variation for sources off beam center (Schneider 1989).
The 22 cm feed has a beam size of $3.3'$ (FWHM), so the gain would be 
about 4$\times$ smaller than its value at beam center for a source in the
worst possible position on the grid. However, for extended sources
the gain variations are smaller, and sources between beams might be
detected at neighboring grid points, so the sensitivity probably varies
by less than a factor of three anywhere in the survey region.

At each point a one minute integration was made over a bandwidth
corresponding to redshifts of $100<cz_{hel}<8340$ km s$^{-1}$. 
This nominally corresponds to distances between 1.3 and 111 Mpc for a
Hubble Constant of 75 km s$^{-1}$ Mpc$^{-1}$ (assumed throughout this paper)
and a total survey volume of $\sim$7630 Mpc$^3$.
We tuned the 22 cm feed to its highest tunable frequency, corresponding
to a redshift of $cz_{hel}=5300$ km s$^{-1}$, for these
observations. The sensitivity rolled off to 81\% of its maximum at
the high-redshift end of the bandpass, and 56\% at the low-redshift end.

During a given observing run, points on the hexagonal grid spaced by
$1^m 12^s$ were observed in sequences of 15. The offset nearly
equaled the integration plus slew time between points so that the telescope
remained in a nearly fixed configuration. Because sources are uncommon, the
other 14 observations in a sequence could be combined to make a low
noise ``off'' (calibration) scan,
and the noise in the final spectrum comes almost entirely
from the ``on'' scan alone. This mapping method is about four times
more efficient than the standard position-switching procedure for a single
point, which would require 2 min ``on'' plus 2 min ``off'' to reach a similar
noise level.

The autocorrelation spectrometer channels were spaced by 39 kHz
($\sim8.2$km s$^{-1}$). After Hanning smoothing the resolution
was $\sim16.4$km s$^{-1}$ giving us undiminished sensitivity to sources with
line widths as narrow as $\sim32$km s$^{-1}$, which is about as narrow
a velocity spread as any dwarf galaxy is known to have.
After carrying out a polynomial fit to channels away from interference or
signals, the noise was typically 1.7 mJy. With corrections for the
frequency response of the feed, the average noise across the band
was about 2.0 mJy.

\subsection{Signal Identification}

Both software techniques and visual examination were used to identify
possible signals in each spectrum. The software techniques included a range of
velocity smoothings and a variety of interference identification and rejection
schemes. The dominant problem was the rejection of weak sources of
interference. Strong interference was relatively easy to isolate because it
would normally persist for more than one minute; thus it would appear in
neighboring spectra in time, but these were separated by 4 beam spacings
($\sim16'$) on the sky and unlikely to be real sources. 
We also identified interference by comparing the left
and right circular polarizations; most man-made signals will appear
polarized while HI emission does not.

Weak interference was much more difficult to isolate because polarization
might be indistinguishable from noise variations, and it might only exceed
the search threshold level in an isolated spectrum.
In the end, the by-eye search proved to be the more effective method for
identifying weak signals---the software methods could be adjusted to identify
weak signals, but only at the expense of introducing very large numbers of
spurious sources. We suspect this is because of a variety of
subtle visual cues that are difficult to design into software recognition, like
artifacts of baseline removal, experience of past patterns of interference, or
knowledge of the probable shape of HI profiles. Of course, these same cues
represent potential biases against non-standard sources of HI emission, and
this should be kept in mind when trying to understand the completeness of
the survey.

Overall, we identified 230 positions by eye or by software algorithms
with suspected emission, and these were re-observed with standard
``on--off'' integrations, except for cataloged sources that 
had already been detected at our position and redshift. (Note, though, that
information about previous detections was not consulted before carrying
out our searches.)
For the suspected signals that we did reobserve, our confirmation rate was
about 30\%.
Genuine HI signals were confirmed at 101 of the 14,130 survey
positions. These correspond to only 75 distinct sources since some sources
were detected at neighboring positions, and at some positions more than one
HI source was evident.
We also dropped one galaxy (UGC 1551) from the final sample even though it
generated a genuine HI signal because it was outside of the survey area 
and was detected through a sidelobe in an observation along our
northernmost row.

In order to determine their fluxes and positions more precisely,
all 75 of the HI-selected sources were mapped in more detail.
We designed an observing pattern to examine a central point and six
equally-spaced surrounding positions, 2.3$'$ away from each other,
for one minute each.
A single ``off'' position was tracked for 5 min at the average
configuration of the telescope during the seven ``on'' scans.
For each source, we fit a simple model to the integrated
fluxes at the seven positions of the hexagonal maps.
We treated each as an elliptical-shaped Gaussian distribution of
gas with an unknown size, position, and orientation
on the sky and model the beam as a Gaussian with a FWHM of 3.3$'$.

The HI source number and best-fit coordinates of the HI are listed
in Table \ref{tbl-1},
columns (1) and (2).\footnote{The numbering and identification of the
sources in this paper supersede earlier listings based on a 
preliminary analysis (Spitzak 1996; Schneider 1996).}
Optical images (described in \S 3.2) centered on the HI positions are
shown in Fig.~\ref{images}, except for a few cases where the 
HI position is marked by a white $+$ symbol.
Based on comparisons to the coordinates of unambiguous optical
counterparts, we found that the HI positions were systematically too far
south by 21$''$.  This was consistent along the entire slice, so we
have adjusted the declinations accordingly.
The resulting centroid positions have an rms scatter of $17''$
in each coordinate relative to the optical positions, which is
the approximate pointing accuracy of the Arecibo telescope.

At several mapped positions it was clear that the emission came from more
than one source, based on variations of the velocity ranges and profile
shapes over the area of the small hexagonal maps.
In these cases (\#20+\#21, \#22+\#23, \#37+\#38, and \#59+\#60),
we assigned the portion of the flux in each hexagonal map position according
to its suspected source. For \#20+\#21 and \#59+\#60 the distinctions
were sufficiently clear that we could fit to the HI of each source separately,
and the individual coordinates are listed in Table \ref{tbl-1}.  For the other
two pairs, the separation was less clear, but we have made our best estimates
of the individual HI parameters. We believe that all
of these sources would have been individually detected if they had been
isolated, and we therefore list them as separate sources in our sample.
In addition, source \#40 was along the northern limit of our survey,
so that we did not have a good initial constraint on its position, and
there seem to be some weak HI signals from nearby companions that
further complicate interpretation of the small hexagonal maps.
As a result we were unable to make a good estimate of its position.
Because of confusion, some measurements of these sources are less
certain, which we note by $\sim$ symbols before entries in the table.

For the interacting pair \#22+\#23 (UGC 12914+5), we consulted the
VLA HI data from Condon et al.~(1993) to estimate the HI properties
of each galaxy.  We interpret their data somewhat differently than they do
based on intercomparisons of our Arecibo and their VLA maps. In particular,
we examined our higher resolution spectra for edges that might
correspond to the upper and lower limits of rotation in the two galaxies
as mapped out in the VLA velocity-channel maps. To us it appears that
UGC 12914 has HI at redshifts from 4030 to 4620 km s$^{-1}$ ranging from the
northeast to the southwest. By contrast, the HI near UGC 12915 has
redshifts from 4160 to 4700 km s$^{-1}$ ranging from the southwest to the
northeast. This would indicate a larger velocity difference between the
two galaxies than found by Condon et al. The gas is clearly asymmetrically
distributed around each galaxy, with a large amount probably associated with
a tidal tail extending south of UGC 12914 and a weaker excess extending 
northwest of UGC 12915. Our total flux of 15.75 Jy km s$^{-1}$ is in good
agreement with the VLA value of 15.3 Jy km s$^{-1}$, and based on the VLA
maps we estimate that 60\% of the total
HI emission is associated with UGC 12914 and 40\% with UGC 12915.

The heliocentric redshift, and line widths at 50\% and 20\% of peak are
listed in columns (3)--(5). The shapes of the HI profiles are shown in
Fig.~\ref{images} along
the bottom of each panel; the panels are 1000 km s$^{-1}$ wide.

In column (6) we indicate the ``number of horns,'' $N_h$, in each profile. 
Single-horn profiles are Gaussian or triangular in shape,
and are usually associated with face-on galaxies or dwarfs.
A two-horned profile is the signature of a rotating disk system: because
the disks have a nearly constant rotation speed and are usually seen
at a nearly fixed inclination, the gas clumps kinematically at two
extremes. These profiles normally have steep edges and weaker emission
in the center. We list $N_h=3$ for profiles which show an
extra central peak, or which have the relatively square shape of a
double-horned profile, but in which the central part of the profile is higher
than the horns. These ``triple-horned'' profiles indicate the presence of
a large amount of gas moving at lower speeds along the line of sight than
the gas forming the rotation horns. This may be caused by warped disks, tidal
interactions, or confusion with companions.

For normal double-horned profiles, the 50\% width is the less biased
estimator of galaxy rotation (Corbelli \& Schneider 1997). However,
in 9 instances the 50\% and 20\% widths vary by more than 50 km s$^{-1}$,
suggesting complications that need to be considered more carefully.
For source \#49 the cause appears to be noise.  Sources
\#12 and \#36 have shallow-sloped edges to their profiles and are
clearly interacting with near neighbors based on their optical images;
some outer gas is probably orbiting at a higher inclination than the main disk.
For \#25 and \#31, small shoulders on one side of the profile may actually
be small companions. In all of these sources, the 20\% linewidth would
tend to be overestimated, and the 50\% linewidth remains the better indicator
of galaxy rotation.

The four other sources with linewidth discrepancies (\#1, \#4, \#9, \#57) 
are in our class of ``triple-horned'' profiles. For these galaxies
low-velocity gas has raised the reference flux density against which the
50\% comparison is made, so that it is probably measuring too narrow a line
width. We would therefore recommend using the 20\% line width as a better
indicator of the rotation speed in these and the other triple-horned galaxies.

We list the central flux in column (7)---this is interpolated
to the centroid position of the HI using our model fit to the seven positions
of the small hexagonal map.
We were also able to determine the total integrated HI fluxes 
by summing the seven spectra, weighted so that the
sensitivity is nearly uniform over a $5'$ diameter area.
The total HI flux is given in column (8), and the profiles in the figure
are based on these summed spectra.

\subsection{Other Galaxies in the Search Volume}

Using long ``on--off'' integrations and re-examination of the
survey spectra, we also searched for HI emission
from cataloged galaxies potentially within our search volume.  This added
10 more sources that have 21 cm detections and 4 more with optical redshifts
within the search volume. Two additional sources were also identified during
follow-up HI observations in the region, making a total of 16 non-HI-selected
sources in our search volume.

For six galaxies with optical redshifts (\#82, \#84, \#87--90) we found
features in the survey spectra that probably correspond to their HI emission 
but which we did not originally identify as part of the HI search.
These all fall fairly close to our detection limit, and in the two strongest
sources, interference at a nearby frequency appears to be the reason for their
rejection in both the visual and software searches. The signal from source
\#87 is at least partially confused with the nearby pair \#37+\#38; we include
it as a detection since there appears to be some weak emission in velocity
ranges not covered by \#37+\#38.

Source \#83 was detected during confirmation
observations in the ``off'' scan of a suspected source. Re-examination
of the search spectrum nearest its position shows weak evidence for this
source, but it was again close to our detection limit.
Source \#91 was found during the detailed mapping of source \#68,
which triggered the detection, but
we do not think \#91 would have been detected on its own,
so we include it in the non-HI-selected group.

These non-HI-selected sources provide useful checks on the limits of our
survey's sensitivity, and they provide an interesting comparison sample.
We include the available 21 cm
information for these sources in Table \ref{tbl-1}. Values extrapolated from
observations made at nearby positions are marked with a $\sim$ symbol.

Two close companions to HI-rich galaxies had reported HI detections
that we believe are due to confusion based on our own mapping. IC 1559
(\#85) is just $0.4'$ southeast of NGC 169 (\#36). It has a listed HI
detection in the compilation of Bottinelli et al.~(1990), but we cannot
identify the original source of this value. Our mapping does not show
any evidence of independent HI emission from IC 1559, but higher spatial
resolution is needed to make a definite determination. Similarly,
IC 5242 (\#76) is 2.7$'$ northwest of the HI-rich IC 5243 (\#3). Our map
confirms the doubts of Giovanelli et al.~(1986) that IC 5242 was actually
detected at 21 cm. Both of these sources do have optical redshifts, though,
so they remain part of our non-HI-selected sample.

Finally, we note that several sources with previous reports of HI
emission generated negative results.  We could not confirm the marginal
detections of UGC 12663, CGCG 482-050, or IC 190 reported by
Giovanelli \& Haynes (1989) at redshifts that would have placed them
within our search volume. 
Our spectra had lower rms values than theirs, and we had experienced
interference at the frequency of the reported detection of UGC 12663.
Since we know of no optical redshifts for these sources, we treat them as
sources of unknown redshift and drop them from further consideration.

\subsection{HI Detection Limits}

To aid in understanding the detection statistics of our survey, in
columns (9) and (10) of Table \ref{tbl-1} we list the integrated flux and the
line width at 20\% of peak as determined from the original search
spectra before making any corrections for frequency response or pointing.
The detected flux is affected strongly by the position of the
telescope beam relative to the source; the offset of the HI centroid
from the nearest point on the search grid is given in column (11).

Understanding the way in which survey sensitivity depends on observational
factors is essential for deriving population statistics from such a sample
(Schneider 1997; Schneider et al.~1998).
In principle, a source of a given integrated flux should be harder to
detect if it has a wider profile since noise is added to it in proportion
to the square root of the number of channels
(Schneider 1996; Zwaan et al. 1997).
We plot the detected fluxes against their linewidths in Fig.~\ref{fluxwidth},
and we show a ``5--$\sigma$'' detection limit based on this statistical noise
argument. This is fairly good at delineating the difference between detected
(solid symbols) and undetected (open symbols) sources.  The boundary is
somewhat fuzzy, though, and it appears that wider profiles are somewhat
more difficult to detect than the statistical uncertainties alone would
suggest. We explore this more thoroughly in our analysis of the HI mass
function (Schneider et al.~1998).

\addtocounter{figure}{1}
\begin{figure}[tb]\plotone{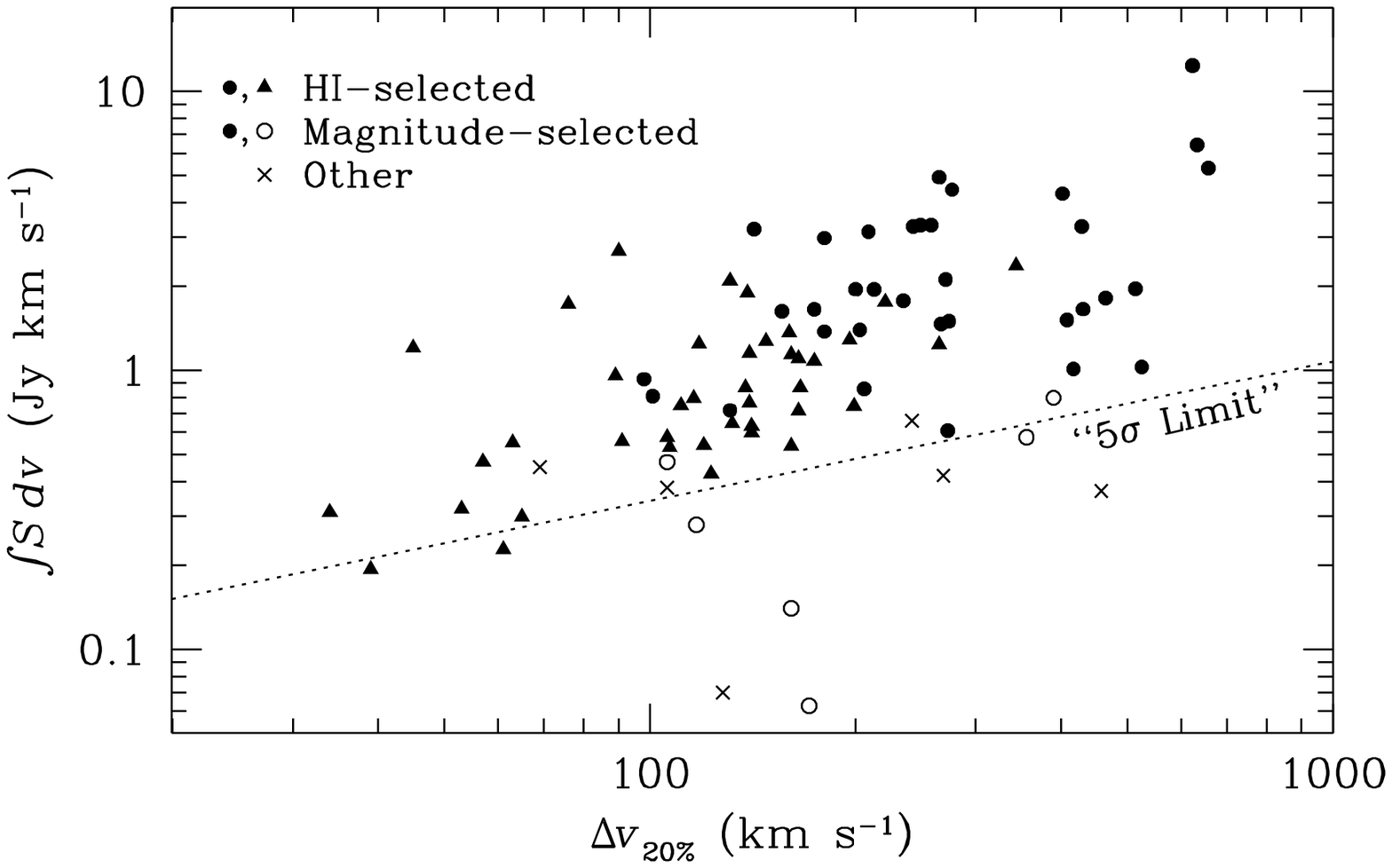}
\caption{Observed HI flux versus line width for sources in the Arecibo
slice. HI-selected sources within our survey are shown by filled
triangles and circles. Open and filled circles indicate
magnitude-selected objects. Sources selected by neither criterion 
are marked by $\times$'s. The line shows a simple
prediction of the 5--$\sigma$ detection limit for HI sources. (See text.)
\label{fluxwidth}
}\end{figure}

Fig.~\ref{fluxwidth} also introduces a notation we shall use throughout
the remainder of this paper. We mark the objects detected as part of our
HI-selected sample by filled symbols, and magnitude-selected sources
(defined as being in any of the magnitude-limited catalogs discussed in
\S 3.1) by circular symbols. Sources belonging to neither
set are marked by $\times$ symbols. The HI-selected sample is split
roughly in half between ``optically-bright'' HI sources (shown as filled
circles) and ``optically-faint'' HI sources  (shown as filled triangles).
This division provides us with the opportunity to examine selection
effects influencing the inclusion of galaxies in different catalogs. 
We will show in \S 4.2 that the optically-bright HI sources are in fact
``star-dominated'' while the optically-faint HI sources are ``gas-dominated.''

\subsection{VLA Observations}

Based on the Arecibo results, several of our objects have
quite extreme properties. We targeted
sources \#14, \#17, and \#75 with the VLA in D-array primarily
to determine precise positions of the HI to aid in their optical
identification.
In addition source \#18 was detected in the field of source \#17.

The observations were 10 min long ``snapshots'' with a
frequency resolution after on-line Hanning smoothing of 97.7 kHz
($\sim$21 km s$^{-1}$). The flux calibration was based on VLA standards,
and for each source a nearby phase calibrator was observed at the same
Doppler-shifted frequency as the HI source.
The images were reduced using standard {\it Astronomical Image Processing
System} ({\it AIPS}) procedures. The synthesized beams were 
nearly circular with $\sim50''$ HPBW, and the rms
noise was $\sim$1.4 mJy per beam.

The VLA results are given in Table \ref{tbl-1a}. Coordinates are listed
in column (2) and marked by white $\times$ symbols in Fig.~\ref{images}.
The offset from the Arecibo position is given in column (3).
The {\it AIPS} Gaussian fits to the source positions quoted errors of
$<1''$, but it seems more reasonable, given the beam size, to suppose
an accuracy of 5--10$''$.
The integrated flux, as a fraction of the Arecibo total flux is given in
column (4), and the deconvolved HI dimensions in column (5).

\begin{figure}[tbp]\plotone{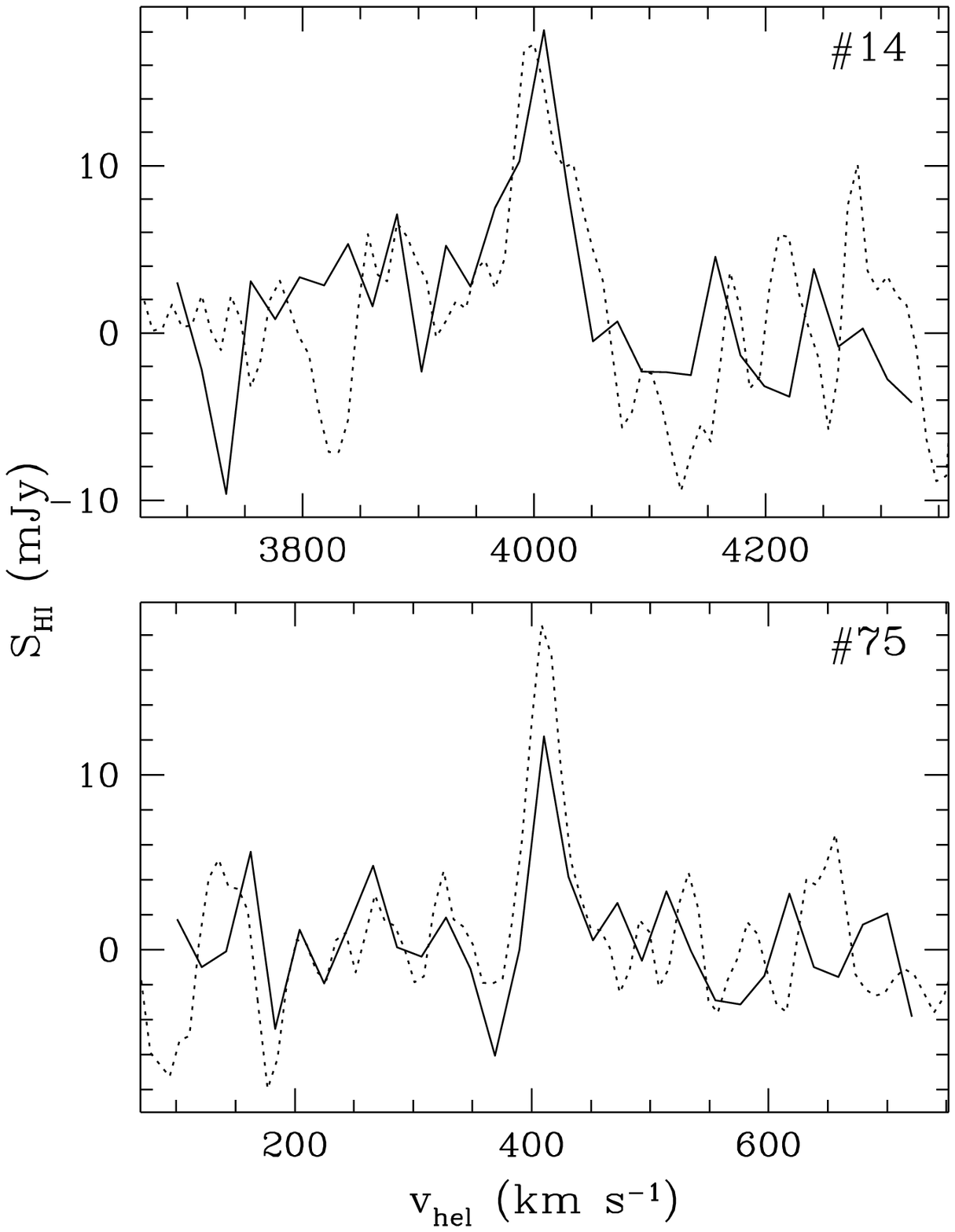}
\caption{Comparison of Arecibo and VLA spectra for two unusual HI sources.
Source \#14 has the lowest surface brightness of any of our sources, while
\#75 has the lowest redshift and lowest mass of any of our objects.
The VLA spectra are shown with a solid line, and the Arecibo spectra with
a dotted line.
\label{VLAspectra}
}\end{figure}

The VLA position for source \#14 indicates it is associated with an
extremely low surface brightness object east of the Arecibo
position.  The optical data presented in \S3 show that this object
has the lowest surface brightness of any of our sources.
The VLA and Arecibo spectra are compared in Fig.~\ref{VLAspectra}.

Source \#17 is over $1'$ south of its Arecibo position. The problem with
the Arecibo position may arise from sidelobe confusion with source \#18,
which is 5$'$ northeast of the VLA position for \#17.This could also explain
the substantially smaller HI flux determined at the VLA.
However, the VLA detection of source \#17 is concentrated almost entirely
over the velocity range of the high-velocity horn seen in the Arecibo profile.
This suggests that the VLA observations may not have been deep enough to
detect more-extended gas associated with the main disk of the optically
edge-on galaxy seen in Fig.~\ref{images}. One possibility is
that there may be extended tidal debris between these galaxies below
our detection threshold at the VLA; this would accord with the apparent
tidal tail extending to the northeast of \#17 seen in the optical image.

Source \#75 is the nearest and the lowest mass object detected in our survey.
Based on its Arecibo position, it had no clear optical counterpart, and,
unfortunately, the VLA position indicates it is 14.5$''$ south and 6.5$''$
west of the bright star in Fig.~\ref{images}.
The VLA and Arecibo spectra are compared in Fig.~\ref{VLAspectra}.
The VLA flux is smaller than the Arecibo flux, although both are weak and
subject to fairly large uncertainties. The Arecibo flux we give in Table
\ref{tbl-1} is the average of three separate small hexagonal maps; these
maps also suggested that the source might be somewhat extended as is also
suggested by the larger total flux than single-beam flux in the Table. The VLA
observations indicate the HI source is point-like based on an AIPS Gaussian
deconvolution of the source from the beam, so if the larger Arecibo flux
is correct, there may be some extended HI emission in the region.

The VLA fluxes and deconvolved HI diameters allow us to estimate an
average face-on surface density of HI in these systems.\footnote{Confusion
and the uncertain geometry of \#17 make it
impossible to estimate an appropriate surface density from these data,
although a simple analysis would suggest a value between \#14 and \#18.}
Source \#14 has an estimated surface density of $2\times10^{20}$ cm$^{-2}$,
which is well below the threshold for star formation of $\sim10^{21}$
cm$^{-2}$ found by Taylor et al.~(1994). Our beam does not allow us
to resolve clumping of the HI, but the smooth optical appearance suggests
that the density of HI may be low throughout the disk of \#14. This is in
contrast to the relatively bright (although also uncataloged) source \#18,
which has a surface density of $9\times10^{20}$ cm$^{-2}$ and which appears
optically to have knots of star formation. Finally, since \#75 is unresolved,
we cannot estimate a mean column density; if we assume it is smaller than
$0.5'$, the surface density would be $>7\times10^{20}$ cm$^{-2}$, which
does not provide us with any firm indication of whether we should expect
to see current star formation.

\section{Optical Observations}

\subsection{Optical Cross-Identification}

After detecting our sources, we initially searched for optical counterparts
by examining the POSS.
Almost all of the sources have at least one likely counterpart within
a small distance of the HI position.
Fig.~\ref{images} shows our own $R$-band CCD images for 
the HI-selected sources (see \S 3.2 for details of this imaging).
Each field is $3'\times3'$, centered on the position of the Arecibo
HI detection, except \#21 and \#40 where the HI positions are noted
by $+$ symbols.

In Table \ref{tbl-2} we give the optical coordinates of the source we
judge the most likely association in column (2).  Generally this
is the brightest object within $1'$, although in regions where
there is more than one candidate we have compared all of the data from
the survey and hexagonal maps, noting profile asymmetries and relative
strengths to identify the most likely association.
This is not a guarantee that the optical and HI sources are the same,
but in all but a few confused regions discussed below, the association
seems nearly certain.

The positions were determined from the Digital Sky Survey (DSS) by
extracting images centered on the HI coordinate, and then
determining the offset to the optical source in pixels from the image center.
The accuracy of these positions is close to the $1.7''$ pixel size of the
DSS images: for 19 sources with high-accuracy measurements listed in the
NASA/IPAC Extragalactic Database (NED), our rms positional differences were
2.3$''$ in $\alpha$ and 1.4$''$ in $\delta$. Compared to the precision
measurements of Klemola, Jones, and Hanson (1987) for nine of our sources
the rms differences are 1.0$''$ in $\alpha$ and 1.1$''$ in $\delta$.
The offset of these objects from the Arecibo HI positions are
given in column (3).

Source \#75 is too close to a bright stellar object seen in the image to
identify it with an obviously extended source. The stellar object is almost
certainly a foreground star, although since this is our lowest redshift and
lowest mass source, it is unresolved in our VLA observations, and it is
the only one so closely aligned with a bright
stellar object, its identification as a star should be confirmed.
Unfortunately, since this source has the latest hour angle of any of our
sources, the CCD observations had to be made at about two airmasses, so
they are not quite as deep as for the other sources. If the source is no
closer to the star than indicated by the VLA position, we can rule out
an extended source with a surface brightness 
brighter than about 25 mag arcsec$^{-2}$, which would be detectable at this
position based on experiments with the image. We also made long
H--$\alpha$ observations of the region, and the H--$\alpha - R$ image,
which cancels out light from the bright star, shows no evidence
of an H--$\alpha$ source in the redshift range of the HI source.

Source \#39 appears to be a clump of several small, faint objects
embedded in an even fainter background; this may be an interacting system.
Source \#57 is probably a composite spectrum of the two objects visible
in Fig.~\ref{images}, but we could not separate the HI signals. The large
line width is probably associated with the edge-on galaxy east
of the HI centroid, and that is the galaxy for which we list optical data.

We searched NED for corresponding cross-identifications, and these
are given in columns (4)--(8) of Table \ref{tbl-2}. Actually, only
37 of our HI-selected sources had been cataloged in NED when the planning
of this survey began, but a number of sources have been subsequently
identified in a variety of surveys. The kinds of sources detected by
various surveys is an interesting study in differing selection effects.
In Table \ref{tbl-2} we divide the identifications into several categories:
magnitude-limited surveys; diameter-limited surveys; emission-line and
ultraviolet excess surveys; and far-infrared surveys.

Column (4) lists cross-identifications with the three major
magnitude-limited galaxy catalogs covering the region:
the NGC/IC, the CGCG, and the MCG (see table notes for references)
in that order of precedence. The NGC/IC sources are fairly
complete down to a photographic magnitude limit of 14, and the CGCG and
MCG extend this to about 15.7, although the limits of the MCG seem much
less uniform. These catalogs identify 35 of the 75 HI-selected sources.

Column (5) lists sources in diameter-limited catalogs. The UGC is
diameter-limited to an angular size $\geq1'$ at the effective surface
brightness limit of the POSS. It picks up only two of the HI-selected
sources missed by the magnitude-limited catalogs; on the other hand, it
misses only seven of the magnitude-limited sources. In addition, a 40$''$
diameter-limited search for edge-on, ``flat'' galaxies (FGC)
detected one more of the HI-selected sources.

Most of the HI-slice region (covering sources 1, 2, 4--7, 27--55, and 71--75)
was examined on the POSS II by J.~Schombert (references LSBC and ESDO in
the Table) in search of dwarf and LSB galaxies to a diameter limit
of 20--$30''$. He detected four more sources, two of which exceed the
1$'$ UGC limit on the POSS II plates, which are approximately 1 mag deeper
than the original POSS.  Over the area covered by plates that Schombert
searched, there are 19 HI sources not in other magnitude- or
diameter-limited catalogs (17 found only in the HI survey), most of which
are LSB objects not included in his lists.

In Column (6) we list cross-identifications from UV-excess and emission-line
surveys.  For sources identified in numbered lists (Mrk, Ark, Kaz),
we give the source number.  In catalogs that list sources
by their truncated coordinates, we note only the catalog name.
These other catalogs are the Universidad Complutense de Madrid survey of
emission-line galaxies (UCM), the Kiso Ultraviolet Galaxy survey (KUG),
and the Hamburg QSO Survey (HS).

The UCM survey covers approximately 36$^\circ$ along various sections of
the HI slice. Of the 44 HI-selected sources in their search regions
(\#\# 3, 4, 8--19, 22--26, 34--58), 7 were detected. The KUG survey covers
the slice at right ascensions earlier than $00^h12^m$, where it detected
8 out of the 27 HI-selected sources. In areas covered by at least one of
these surveys, they detected 4 of 18 sources that were not detected by the
magnitude- or diameter-limited surveys.

In column (7) we note whether the source was detected in the far-infrared
(FIR) by the Infrared Astronomical Satellite (IRAS). If so, the published
60 and 100$\mu$ fluxes are listed (Moshir et al.~1990; Zamorano et al.~1994).
The FIR-detected sources are generally among the brightest of the galaxies,
although there are a few interesting exceptions, including two of the
emission-line-detected sources and the accidentally-detected HI source \#91.
We should note, however, that the accuracy of the FIR source positions is
uncertain enough that some associations may be spurious.

Finally, in column (8) we list morphological classifications given in NED.
The designation ``pec'' indicates that the morphology was identified as
unusual or that the source appears in a catalog of peculiar or interacting
sources (Vorontsov-Velyaminov 1959; Arp 1966; Zwicky 1971).
These peculiar objects are often tidally disturbed galaxies, and they show
a high degree of overlap with the FIR, UV-excess, and emission-line sources.

\subsection{Optical CCD Observations}

To study these objects in more detail, we obtained broad-band optical
images of all but one of the HI-selected sources, along with several of
the non-HI-selected sources. These data were collected using the 0.9 meter
NOAO telescope at Kitt Peak, Arizona in two observing sessions using Tektronix
CCDs with $1024\times1024$ or $2048\times2048$ arrays and a pixel spacing
of 0.60$''$ and 0.68$''$ respectively. Images were obtained using Johnson
$B$, $R$, and $I$ filters, generally with 5--10 min integrations and multiple
exposures for the fainter sources.

The regions around our HI positions shown in Fig.~\ref{images} are
logarithmic stretches of the $R$-band images. All of the images are
shown with the same stretch so that they may be directly compared.
The stretch is designed to keep the central regions from being ``burned
out'' while revealing outer LSB features. This stretch makes some of the
stellar images look larger than the seeing, and it brings out some
artifacts. Linear features rotated a few degrees clockwise from the
cardinal directions are due to diffraction spikes. (Also
note that the center of the subimage that was flattened for these
images was based on telescope coordinates, which is why some do not
extend to the edge of the frame in the figure.)
We did not obtain imaging for source \#11, so we have substituted
estimates for the B-band data of \#11 based on the DSS image. This was
calibrated relative to several other galaxies covering the same surface
brightness range on the same POSS plate. We have attempted to stretch
the POSS image similarly for the figure. Also, because of some technical
problems across part of the $R$-band image of \#10, we display its
$B$-band image in Fig.~\ref{images}.

The optical data were calibrated using Landolt (1992) standards, airmass
corrected, and flattened using standard {\it Image Reduction and Analysis
Facility} ({\it IRAF}) procedures. The seeing was typically about $1.5''$
for the $B$ and $R$ images. Most of our $I$--band data was collected under
poorer seeing of up to $\sim3''$.  Using the {\it STSDAS} package of
{\it IRAF}, elliptical isophotes were fit to the
$B$-band images from the center of each galaxy out to well beyond any
detectable signal at 2 pixel increments of the isophotal radius. Where
the signal became too weak, the ellipses were held to the same center,
position angle, and axis ratio as the last good fit.
The $B$-band data were generally adequate to make isophotal measurements
to $\sim$27 mag arcsec$^{-2}$, and the $R$-band was about 1--2 mag deeper.
Magnitudes at $R$ and $I$ were determined within the isophotal boundaries
determined at $B$.

The optical measurements are listed in Table \ref{tbl-3}. The $B$-band
magnitude measured within the 25 mag arcsec$^{-2}$ elliptical isophote is
given in column (2). This is uncorrected for Galactic extinction, $A_B$,
which we list in column (3) as determined using the COBE-based maps of
Schlegel, Finkbeiner, \& Davis (1998). This new extinction map has much
higher-resolution and indicates a larger average extinction in this region
than the method of Burstein \& Heiles (1982).

The surface brightness at each radius is affected by the inclination of the
galaxy as well as the extinction, and the face-on surface brightness would
be fainter by
$2.5\log(b/a)$ (where $b/a$ is the axis ratio of the ellipse) if the stars
are distributed in a thin disk.  After making the extinction\footnote{We
do not make internal extinction corrections for each galaxy, since these
are relatively uncertain and not clearly applicable to our LSB systems.
Using the RC3 prescription of $A_B=0.7\log[\sec(i)]$, most of the galaxies
would have an internal extinction of $<0.2$, although the most edge-on
systems could have extinctions of $\sim0.5$.} and inclination
corrections, we interpolate to the ``true'' 25 mag arcsec$^{-2}$ isophotal
size and determine a corrected $B^0_{25}$ magnitude which is given in column
(4).  We also estimate the ``total'' extrapolated magnitude in column (5),
based on fits to the light distribution described in \S 3.3 below.

The major axis diameter of the elliptical fit at the corrected 25 mag
arcsec$^{-2}$ isophote is given in column (6). The Holmberg diameter is
given in column (7), and the ``effective'' or ``half-light'' diameter is
given in column (8). This last measurement is the size within which half
of the total light is contained, again based on fits to the light
distribution described in \S 3.3. We list two estimates of the axis
ratio, at the 25 mag arcsec$^{-2}$ and Holmberg isophotes, in columns
(9) and (10), and the position angle at 25 mag arcsec$^{-2}$ in column (11).
The two axis ratio values give some indication of the uncertainty in the
galaxy inclination and whether the outer disk may be warped.

In columns (12) and (13) we list interpolated values of the $(B-R)$ and
$(B-I)$ colors within our $B^0_{25}$ isophote. To account for the wavelength
dependence of the extinction, we assume $A_R=0.47A_B$ and $A_I=0.32A_B$.
We also list the central $(B-R)_0$ color in column (14) determined at the
brightest pixel near the center of the galaxy and extinction corrected as
already described.

\begin{figure}[tbp]\plotone{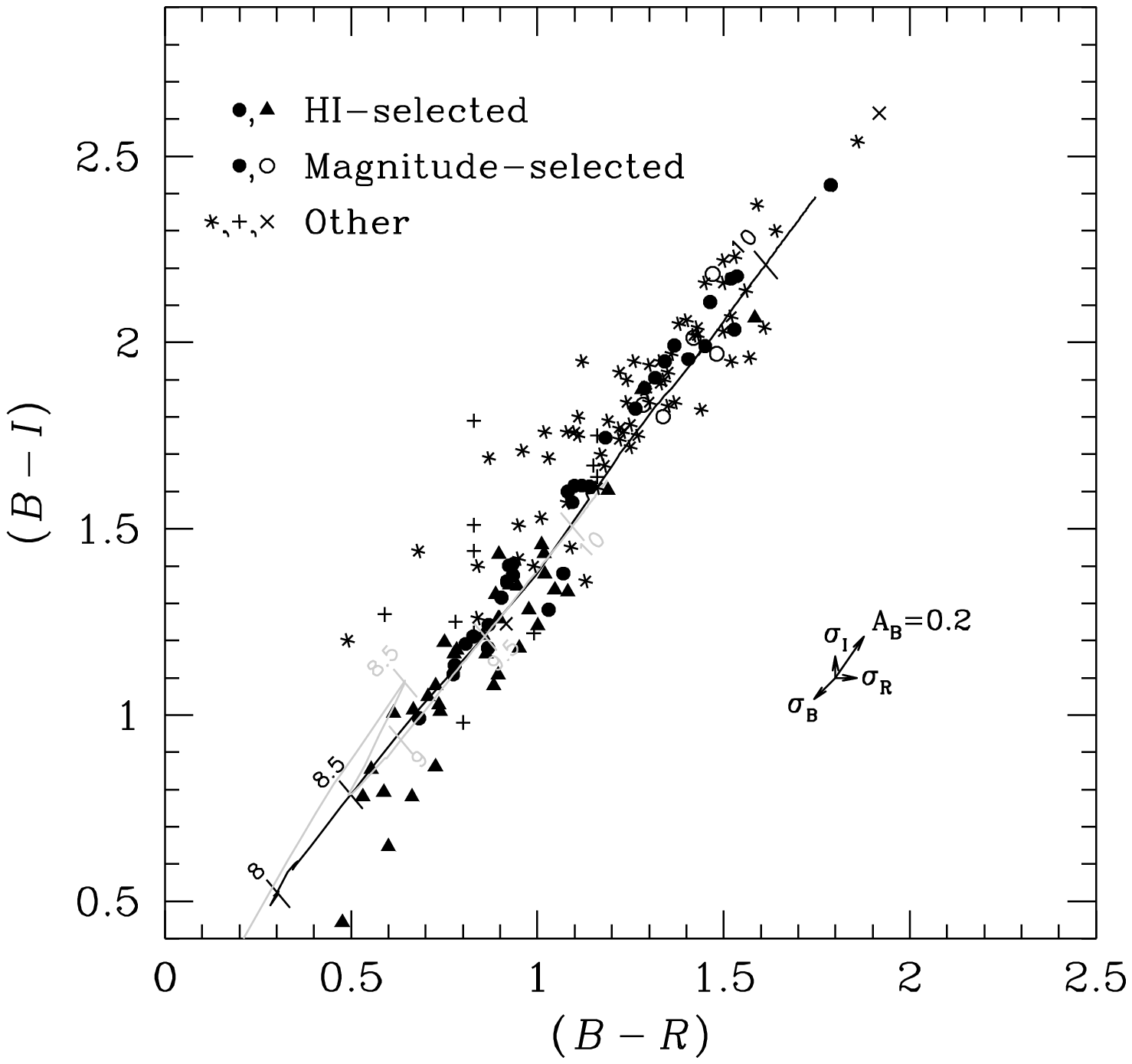}
\caption{Color--color plot of galaxies observed in $B$, $R$, and $I$
in this study. The magnitudes are determined within the 
extinction-corrected $\mu_B=25$ mag arcsec$^{-2}$ isophote.
Filled circles and triangles are HI-selected sources; open and filled circles
are magnitude-selected.
At the right of the figure are shown the effect of the estimated 1$\sigma$
errors in each magnitude and of an internal extinction of $A_B=0.2$ mag.
Data for HI-selected sources of Szomoru et al.~(1994) are shown by $+$ symbols
and data from the diameter-limited sample of de Jong \& van der Kruit (1994) by
$\ast$ symbols.
Black and gray curves show a stellar population synthesis model of the
color evolution of populations with solar and 1/50th solar metallicity
respectively, with log(age in years) marked at half decade intervals.
\label{briplot}
}\end{figure}

We checked our $B$--band data against 18 aperture-photometry measurements
for six galaxies in the catalog of Longo \& de Vaucouleurs (1983). We
determined our flux within the same circular apertures as listed in the
catalog, and find a mean difference of $0.00\pm0.04$ mag from our values after
throwing out one discrepant point (which also disagreed with other
measurements in their catalog). We believe our errors contribute only a
small part of the rms scatter of 0.17 mag based on the small dispersion
in our $B-I$ versus $B-R$ colors (Fig.~\ref{briplot}).

The dispersion about a best-fit linear relationship of
$(B-I) = 1.36\times(B-R) + 0.03$ is only 0.097 mag. If we assume the
$B$, $R$, and $I$ magnitudes each have the same intrinsic dispersion,
we would recover this relationship if the true relationship was
$(B-I) =1.41\times(B-R)-0.03$ and a 1--$\sigma$
uncertainty in each magnitude of 0.056. This assumes no intrinsic scatter
in the relationship, so the actual uncertainty is presumably smaller.

This figure also shows the color data for HI-selected galaxies studied by
Szomoru et al.~(1994), and the data from de Jong \& van der Kruit (1994) for
a complete, diameter-limited sample of galaxies ranging from type Sa to Irr.
Their quoted errors appear to account for the scatter relative to our
empirical relationship, although there is a slight indication that the
$(B-I)$ values might be slightly redder than ours toward the blue end of
the curve.  None of their sources reach the extreme blue colors of our ten
bluest objects, which have mean colors of $(B-R)=0.63$ and $(R-I)=0.26$.

\subsection{Surface Brightness Structure}

In the last four columns of Table \ref{tbl-3} we describe the surface
brightness structure of our objects. Column (15) gives the
central surface brightness as measured in the central brightest pixel of the
$B$-band image, corrected for extinction. The remaining columns of the
table describe our best fits to the $B$-band surface brightness distribution
with combinations of disk and bulge components.

The fits are based on elliptical isophotal fits to the surface brightness
data (corrected for extinction and inclination).
As already noted, the inclination correction is only appropriate
if the stars are in a thin disk, which 
may be inappropriate for dwarf, irregular, and elliptical
systems as well as the central bulges of disk systems. We could have
attempted to handle some galaxies and some regions within galaxies
differently, but this would have introduced a degree of arbitrariness we
preferred to avoid. In any case, we made fits with and without inclination
corrections, and while individual galaxies' parameters were altered
slightly, the types of fits and overall sample behavior were unaffected.

\begin{figure}[tbp]
\epsscale{0.6}
\plotone{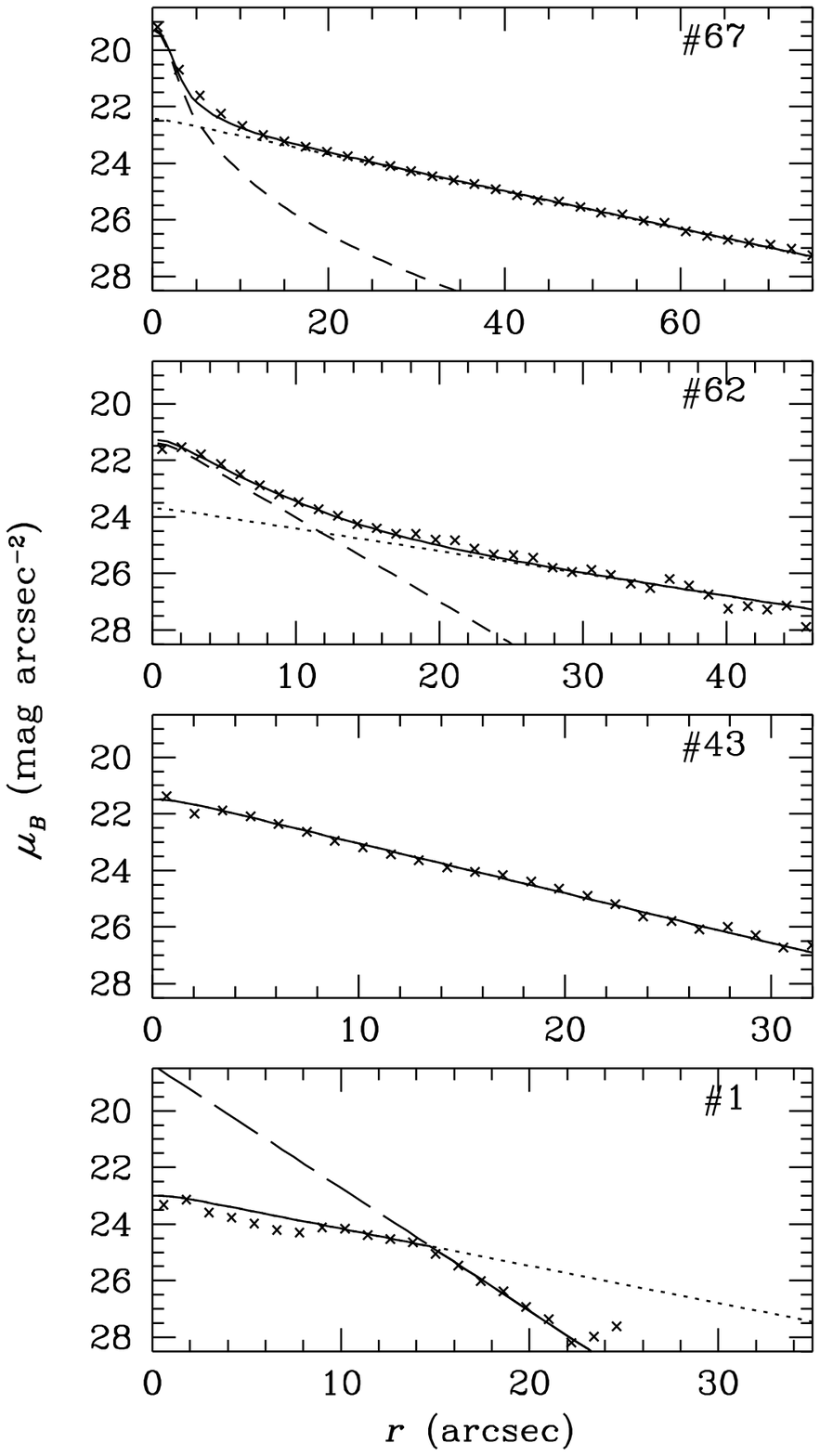}
\caption{Radial $B$-band profiles of four of the HI-selected galaxies,
illustrating various kinds of fits. Source \#67 has a traditional combination
of an exponential disk (dotted line) and $r^{1/4}$-law bulge, while source
\#62 has an exponential bulge (bulges shown by dashed line). Source
\#43 shows only an exponential disk, and source \#1 has an exponential disk
that steepens in the outer portion (long dashed line).
\label{4profiles}
}\end{figure}

We initially attempted to fit the radial profiles by a traditional
combination of an exponential disk with an $r^(1/4)$-law bulge, 
exemplified by source \#67 in Fig.~\ref{4profiles}. As has been noted by
Andredakis \& Sanders (1994), for many late-type galaxies this does not
yield as good a fit as using an exponential fit to the bulge.
An $r^{1/4}$-law fit is forced to have a shallow peak to match the central
light distribution in these galaxies, which in turn forces the outer
portions of the fit to be too high.
Source \#62 in Fig.~\ref{4profiles} is an example of a source which
is better fit by an exponential bulge.
In column (16) of Table \ref{tbl-3} we list what fraction of the total light
from the galaxy is contributed by the bulge component ($f_B$). This fraction
was fairly consistent whichever type of bulge was fit to a particular galaxy.
The value is marked by a $q$ to indicate a ``quarter-law'' ($r^{1/4}$) fit
or by an $e$ to indicate an exponential fit.
The scale length and surface
brightness parameters of the bulge fits depended sensitively on our
handling of the axis-ratio corrections; we do not list them here.

\epsscale{1}

The disk region was fit by an exponential law:
$$\mu(r)=\mu_d + 2.5 {r\over r_d}\log e\ ,$$
where $\mu_d$ is the projected central surface brightness of the disk
component, and $r_d$ is its scale length. This was performed iteratively
with the $r^{1/4}$ or exponential bulge to yield a least squares best fit. 
The exponential disk parameters are listed in columns (17) and (18) of
Table \ref{tbl-3}.

Ten galaxies were best fit by an $r^{1/4}$--law profile without an
exponential disk, and almost all of these had ``bumps'' in their
radial profiles. These bumps also tended to occur in regions where there
were significant changes in the estimated axis ratio, and tended to be
weaker in the inclination-corrected fits, suggesting they may be part of
a disk component that our simple distributions could not model.

For almost half (34 of 75) of the HI-selected galaxies {\it no} bulge
component is evident. This is illustrated by source \#43 in
Fig.~\ref{4profiles}. As can be seen from the values of $f_B$ in
Table \ref{tbl-3},
we fit bulges which contributed as little as 1\% of the total galaxy light,
and these are fairly obvious. We therefore assume any bulge component in
the ``bulgeless'' galaxies contributes $\lesssim$1\%. Note that even
though there is no bulge component, $\mu_d$ does not necessarily match the
observed central surface brightness. This is because $\mu_d$ is fit to the
inclination-corrected surface brightnesses, which can make it fainter than
the observed value, and it is corrected for seeing, which can make it brighter.

Finally, we noticed that the surface brightness profiles appear to turn
over to a steeper slope in several galaxies. This behavior, illustrated by
source \#1 in the figure, cannot be modeled by a sum of components, and
instead represents a cut-off to the fits found interior to it. The portion
of the total light coming from these steep outer exponential disks ranged
from 10 to 40\% of the whole galaxy. This behavior occurs occasionally in
all the various combinations of bulge and disk fits to the profiles, and
has been noted in other studies of LSB galaxies (Davies, Phillipps, \&
Disney 1990; R\"onnback \& Bergvall 1994; Vennik et al.~1996). For
the purpose of estimating the bulge fraction in column (13), we count this
exponential cut-off region as part of the disk contribution. One possibility
is that this change in scale length might be caused by imperfect modeling
of changes in the galaxy inclination or position angle. For those
interested in following up on these systems, we list their cut-off parameters
in Table \ref{tbl-4}. The fraction of the total light in the cut-off region
$f_x$ is listed in column (2), the surface brightness $\mu_x$ where the
turnover begins in column (3), and the scale length $r_x$ of the
exponential cut-off in column (4).

\section{Relationships between Galaxy Properties}

The sample of objects found in this HI survey has a number of properties 
that distinguish it from an optically-selected sample.  We highlight the
properties of our sample here, and discuss general implications for the
overall galaxy population.

\subsection{Spatial Distribution}

We note first that most of the objects we detected at very low redshifts
were not previously detected, suggesting that that the census of objects at
very low redshifts is far from complete. Out to $cz_{hel}=3000$ km s$^{-1}$,
nine sources were found in the HI survey, none of which is in the magnitude
limited surveys, and only three of which were found by diameter-limited
surveys. Even within 1000 km s$^{-1}$, the HI survey reveals three 
objects not found in any other surveys. These are dwarf galaxies,
but their numbers over such a small area and with such a limited depth of
sensitivity implies a very large number density. The implications of these
sources for the galaxy mass function are discussed by Schneider
et al.~(1998).

\begin{figure}[tb]\plotone{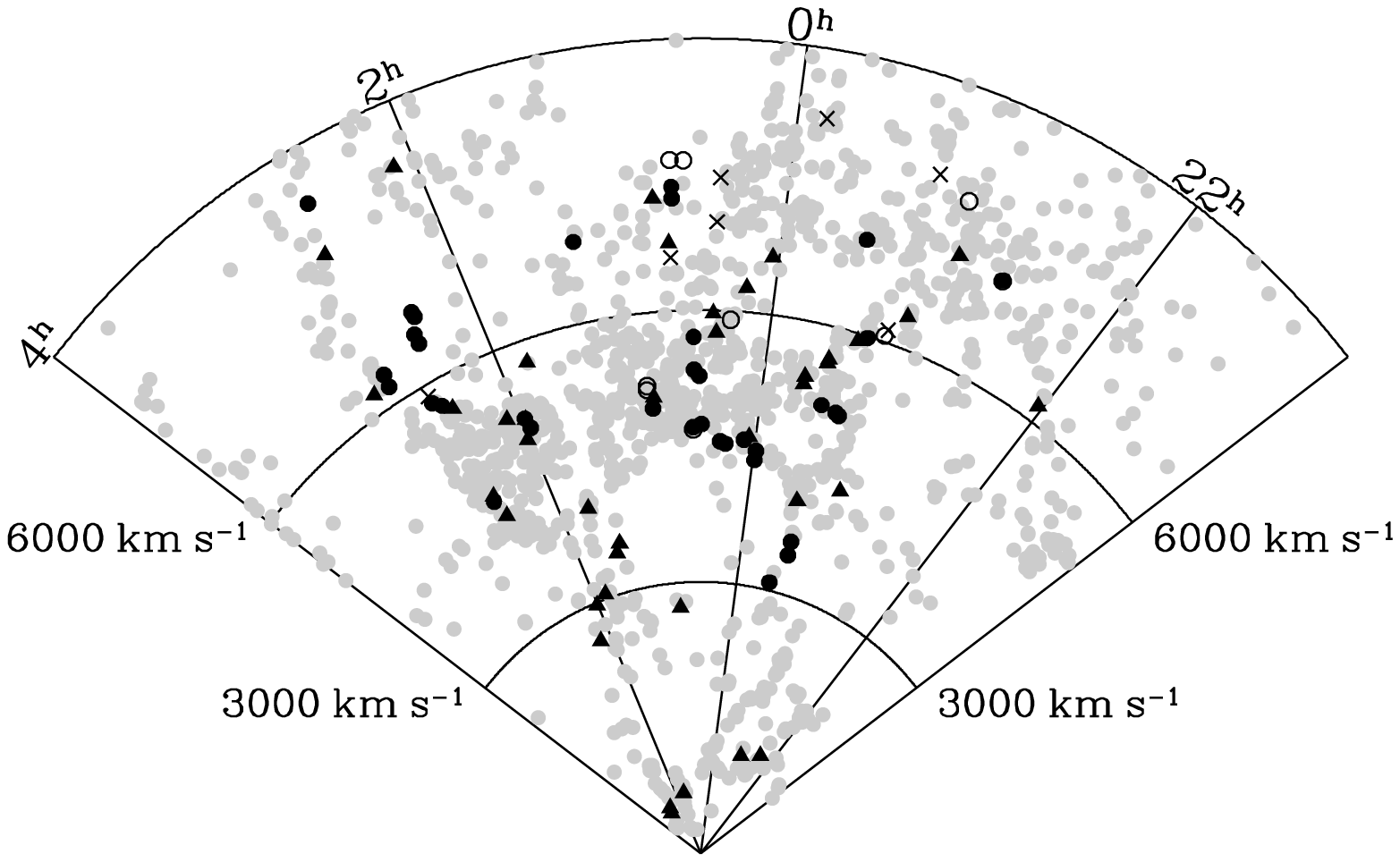}
\caption{Location of slice objects relative to the local large scale structure.
Redshifts of sources are plotted as a function of R.A. Locations of RC3
galaxies north and south of the search region between
$13^\circ<\delta<34^\circ$ are shown in gray. Objects
detected in the slice survey are shown in black:
filled circles and triangles are HI-selected sources; open and filled circles
are magnitude-selected; $\times$'s are neither.
\label{slice}
}\end{figure}

The positions of the Arecibo Slice galaxies relative to the large
scale structure in the vicinity is shown in Fig.~\ref{slice}. Positions of
RC3 (de Vaucouleurs et al.~1991) galaxies within 10$^\circ$ of the slice
(but excluding the slice itself)
are plotted in gray. All of the galaxies, whether they are from our
optically-faint subset or they are LSB objects, basically follow the
same large scale structure traced out by the optical surveys, as has been
noted previously (for example,
Mo, McGaugh, \& Bothun 1994; Thuan, Gott, \& Schneider 1987).

One cautionary note should be made before drawing inferences about
populations in voids: this is a redshift-space diagram, and patterns of
large scale flow around the Pisces-Perseus supercluster and the Local
supercluster may create the appearance of voids where none exist (see
Praton \& Schneider 1994). For example, if the ``bubble'' in the center
of our slice diagram is delineated by velocity caustics of galaxies flowing
toward infall centers in the supercluster, then this region should appear
empty in redshift-space for any population of objects because it represents
only a small volume of real space.

\subsection{Optical versus HI Selection}

\begin{figure}[tbp]
\epsscale{0.4}
\plotone{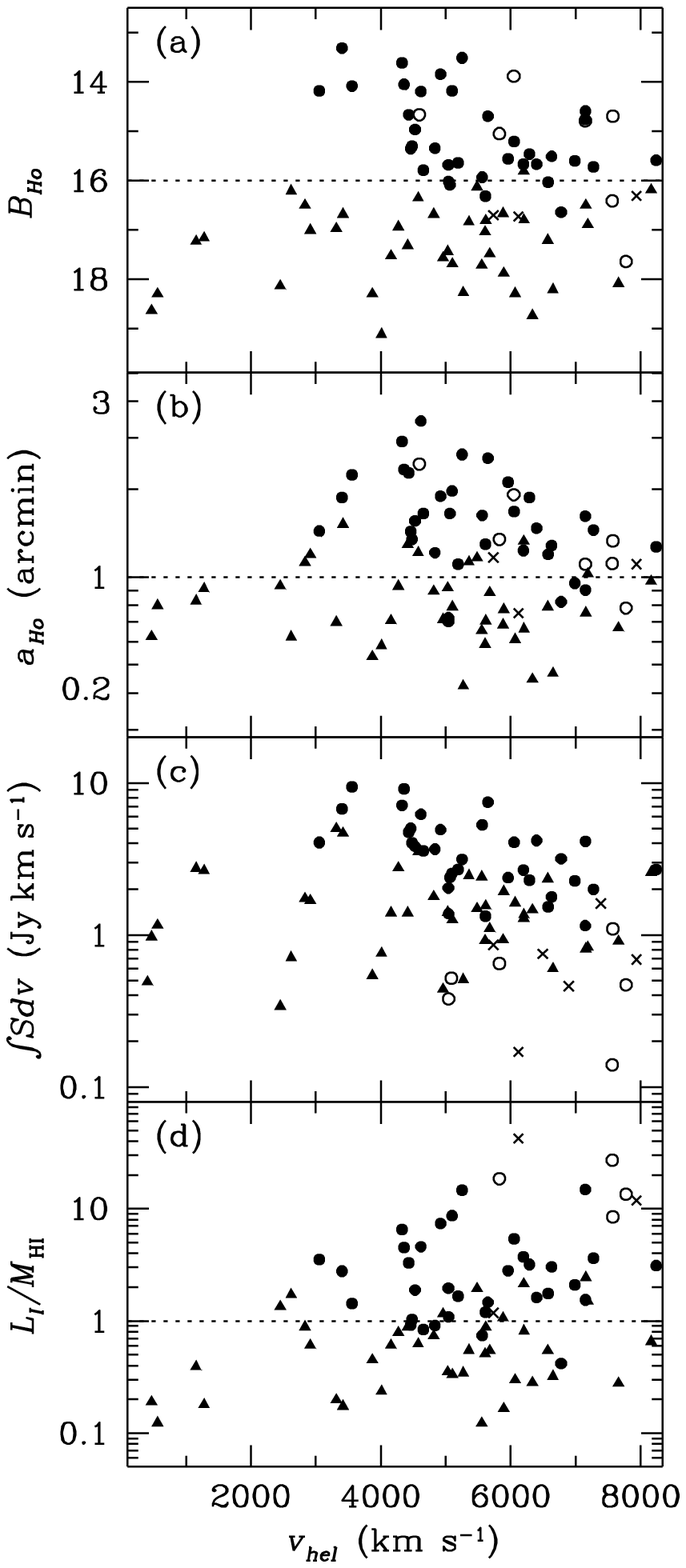}
\caption{Observed properties of sample galaxies as a function of redshift.
Filled circles and triangles are HI-selected sources; open and filled circles
are magnitude-selected; $\times$'s are neither.
Panels (a) and (b) show the observed (uncorrected) Holmberg blue magnitude and
major diameter. Panel (c) shows the HI flux, and panel (d) gives the
ratio, in solar units, of the blue luminosity to the HI mass.
Lines indicate the approximate break between the optically-bright and -faint
halves of the HI-selected sample.
\label{selection}
}\end{figure}

In order to understand the selection effects of different survey approaches,
we plot in Fig.~\ref{selection} various measured quantities as a function
of the galaxy redshift. We again use the notation from Fig.~\ref{fluxwidth},
in which circles represent magnitude-selected sources and solid symbols
HI-selected sources.
Panel (a) shows that the galaxies from the magnitude-limited catalogs
are brighter than 16 mag (marked by a dotted line) with only a few exceptions.
This is as expected, although since these catalogs implicitly depend
on surface brightness and morphological identification, there is at least
the potential for some LSB galaxies with brighter total magnitude
to be missed. We find none such in our HI-selected sample.

\epsscale{1}

In panel (b) we plot the galaxies' Holmberg diameters. Again, the sample
separates fairly neatly into optically-bright and -faint halves at
around 1 arcmin. This accords with the strong overlap between the
magnitude- and diameter-limited samples noted in \S 3.1.
The degree of correlation between the optical
flux and surface area is also shown by the relatively small scatter
in the ratio of these quantities: the standard deviation of
log(optical flux/area) is 0.28 among the HI-selected objects.

In panel (c) we show the HI fluxes of the sources. It is clear that we
cannot draw a simple boundary dividing optically-bright and -faint
sources in this plot.
The standard deviation of log(optical flux/HI flux) is 0.43.
This lack of correlation of the HI and optical
properties is somewhat surprising in light of previous work showing that
the HI content is well-correlated with the optical disk size,
independent of morphological type (Haynes \& Giovanelli 1984). 
A large part of the difference comes from the non-HI-selected sources
which have a much lower mean HI surface density
(measured by the total HI flux and the Holmberg radius), but there
is much more intermingling of optically-bright and -faint sources
than is seen in Fig.~\ref{selection}(b).
If we restrict ourselves to the magnitude-limited subset of our
HI-selected sample, the average HI surface density properties are similar
to the Haynes \& Giovanelli sample, as might be expected since they were
observing optically-selected objects.

Finally, panel (d) shows the ratio of total $I$-band
luminosity to total HI mass. Since both depend on $d^2$, we can find
the distance-independent ``star-to-gas'' ratio of the two:
$${L_I\over M_{HI}}=4.24\times10^4 10^{-0.4(B_T^0-4.02)}/\int S\,dv\ ,$$
where the result is in solar units of mass and luminosity for an
absolute magnitude of the Sun at $I$ of 4.02.

We select the $I$-band luminosity
because it is much less dependent on stellar evolution than $B$-band.
McGaugh \& de Blok (1997) suggest that the mass-to-light conversion from
$I$-band luminosities to total stellar masses is nearly uniform at a
factor of 1.2 for a wide range of stellar
population ages. We find similar values based on a variety of models
of the star-formation history using the population synthesis models of
Bruzual \& Charlot (1993), although for the very youngest blue populations
the stellar mass-to-light ratio may be much less than 1, and for extremely old
populations (which have had no star formation after an initial burst) the
ratio may be as large as 3--4.  To determine the total gas mass, the HI
mass must be corrected by a similar factor of 1.3 to account for primordial
helium. In addition there is potentially a large contribution from molecular
gas, although this is probably larger in the same redder, earlier-type
galaxies in which the stellar mass correction is larger (Young \& Knezek 1989).
All things considered, the $L_I/M_{HI}$ value is probably a reasonable
estimator of the true star-to-gas mass ratio.

The optically-bright sources divide fairly cleanly from the rest of
the sources near a star-to-gas mass ratio of 1, indicating that there is
a physical basis for the distinction between the optically-bright
and -faint subsets of our HI sample than just their apparent magnitudes.
Only 2 of the optically-bright HI sources have a star-to-gas ratio $<0.8$,
and only 3 of the optically-faint HI sources have a star-to-gas ratio $>1.2$.
Thus the magnitude-limited sources are ``star-dominated,'' with an
average of $\sim2.7\times$ more mass in stars than gas, reaching
$>10\times$ among the non-HI-selected objects.
The optically-faint HI objects are ``gas-dominated,'' with an average
gas mass $\sim2.5\times$ the stellar mass, and reaching a factor
$>10$ in the most most extreme case.

Some of the gas-dominated sources were found by other types of surveys, but
these are not our most extreme objects. Seven of them were identified in
diameter-limited and LSB surveys, but the most extreme of these (curiously
a UGC object, \#45) has only the sixth smallest star-to-gas ratio overall.
These 7 diameter-selected sources have fairly typical blue colors of the
gas-dominated subset.
The bluest of these \#2 (LSB F533-01) is only 12th bluest object overall.
These surveys do identify several of our lowest surface brightness objects,
including those with the 3rd, 5th, and 6th faintest $\mu_{1/2}$ values.

Four of the optically-faint HI sources were identified in UV-excess and
emission-line surveys. These objects have colors and surface brightnesses
that are typical of the star-dominated subset, and $L_I/M_{HI}$ ratios
that are borderline between the subsets.  The only significant exception
is \#57, which has the 11th bluest $(B-R)$ color. This object is also
the only gas-dominated source detected by IRAS. It appears to be 
interacting with a small neighbor as we discussed in \S 3.1, and its
nucleus appears to be off center. Two other of these sources also have
off-center nuclei, suggesting that interactions may be responsible for
these sources being detected by alternative methods.

\subsection{Biases in Optical versus HI Selection}

The HI selection criteria diverge in a
fundamental way from the optical selection criteria.
Consider that the HI selection identifies about twice as many sources 
as are found down to a magnitude limit of 16 in the same volume. 
To double the size of a magnitude-limited sample within this volume would
require increasing the magnitude limit by only about 0.8 mag. However,
to catch all but the faintest HI source based on their starlight would
require a survey 4 mag deeper, with about 30 times more sources in this
volume. Further, since optical surveys do not distinguish sources by
redshift, a magnitude-limited survey 4 mag deeper would have about
250$\times$ more sources that would have to be identified and measured
in order to detect these HI-selected sources.

The converse would of course be true if we were using HI to track
down optically bright galaxies. For example, most E and S0 galaxies
would be much easier to identify by optical search techniques than to
find them by the signal from the small amount of HI present. The
implications of this obvious difference are a little more subtle, however.

If we were to define the HI properties of S0 galaxies from our
HI-selected sample, we might erroneously conclude that their HI content 
is fairly large. The four S0 galaxies in our HI-selected sample have
an HI emission about $25\%$ of their $B$-band emission (in solar units).
The one identified S0 galaxy among the non-HI-selected objects has an
upper limit of $<1\%$. Obviously, HI-rich S0 galaxies are over-represented
when we select sources by their HI content. The reverse will also
occur: the HI content of galaxies will be underestimated in
optically-selected samples.

This is akin to a Malmquist bias---flux-limited samples favor sources with
intrinsically larger luminosities. When comparing two 
properties like HI and optical emission, this has the effect of exaggerating
the level of the property being used for the flux limit. The
weaker the correlation between the quantities, the bigger the effect,
so it is important to determine HI properties of galaxies from HI-selected
samples.

\subsection{Surface Brightness Behavior}

The distribution of light varies quite strongly
between the star-dominated and gas-dominated galaxies. We quantify
this here by a closer examination of the radial surface brightness
distributions. We shall also show that these effects do not simply
result from differences due to galaxy mass or luminosity, but appear
to indicate a genuine difference in character of HI-selected galaxies.

The radial fits to the galaxies' light distributions show a wide range of
properties that vary from no bulge component to no disk component, and
exponential or $r^{1/4}$-law bulges (see \S 3.3). These properties
correlate strongly with the HI characteristics of the galaxies.
Dividing the galaxies into the optically-faint and optically-bright halves
of our HI-selected sample and the other magnitude-selected objects,
we list the number of galaxies having no bulge, or an exponential
or $r^{1/4}$-law bulge below:
\begin{center}
\begin{tabular}{lccc}
 & pure disk & exp bulge & $r^{1/4}$-law bulge \\
\cline{2-4} \\
Optically-faint  & 29 & 11 & \phn0 \\
Optically-bright & \phn5 & 19 & 11 \\
Non-HI-selected  & \phn1 & \phn2 & \phn4 \\
\end{tabular}
\end{center}
\noindent
The optically-faint subsample is much more likely to have no bulge at all, and
none have $r^{1/4}$-law bulges. The situation is nearly reversed with
the non-HI-selected galaxies, and the other magnitude-selected sources
have a similarly strong tendency to have bulges, but more likely for
these to be exponential bulges.
Thus 73\% of optically-faint sources have pure
disks, while 86\% of magnitude-limited sources have a bulge component.

Another way of quantifying this comparison is to look at the fraction of
light contained in the bulge component for the different subsets. For the 
40 optically-faint HI sources the bulge contributes on average 5\% of the light,
but for the 35 optically-bright HI sources the contribution rises
to 29\%, and for the 7 magnitude-selected, but non-HI-selected sources
the contribution climbs to 61\%.

Since the fraction of light from the bulge 
is correlated with galaxies' mass and luminosity,
it should be asked whether the trends seen here simply reflect that the HI
sources tend to have lower luminosities and masses. We have avoided bringing
in distance-dependent quantities in this paper,
but to clarify this point we consider objects
within a narrower range of mass and luminosity in which the different
sample subsets are well-represented. For the sources
with dynamical masses estimated between $10^{10}$ and $10^{11}M\sun$,
the bulge fractions are 6\%, 29\%, and 50\% for the optically-faint,
optically-bright, and non-HI-selected subsets respectively.
Very similar numbers are found when selecting
sources with HI masses between $10^{9}$ and $10^{10}M\sun$:
the corresponding bulge fractions are 9\%, 20\%, and 67\%.

What remains clear even when comparing mass and luminosity ranges in which
the subsamples fully overlap is that the HI-selected sources have distinctly
different disk/bulge properties than the optically-selected sources. Still,
this difference might be attributed to morphological-type variations, but
these are often subjective and vague. What would be preferable is some type
of quantitative, distance-independent measure of galaxy properties that
relates relative stellar and gas content of galaxies like the $L_I/M_{HI}$
ratio, which we examine next.

\begin{figure}[tb]\plotone{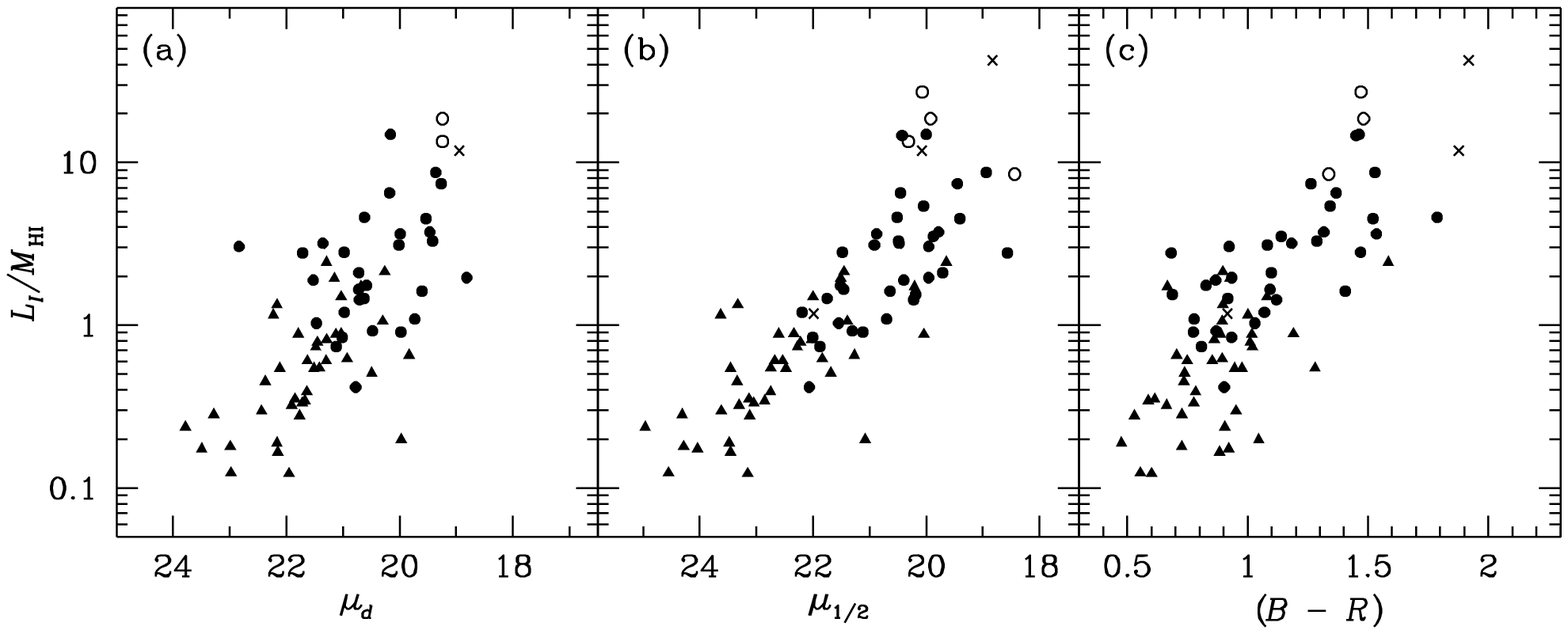}
\caption{Ratio of $I$-band and HI emission plotted against several different
quantities: (a) the disk central surface brightness; (b) the mean
Filled circles and triangles are HI-selected sources; open and filled circles
are magnitude-selected; $\times$'s are neither.
$I$-band surface brightness; (c) the $(B-R)$ color.
\label{sbcorr}
}\end{figure}

\subsection{Surface Brightness Correlations}

McGaugh \& de Blok (1997) have noted a strong correlation between the
star-to-gas mass ratio and the disk central surface brightness. Our
HI-selected sample shows a weaker correlation between these quantities,
as shown in Fig.~\ref{sbcorr}(a).  We find the correlation coefficient
$\rho=0.65$ between $L_I/M_{HI}$ and $\mu_d$, and only $\rho=0.49$ using
the $B$-band data, which is less than the value of $\rho=0.73$
found by McGaugh \& de Blok.  We suspect the stronger correlation they
found may be because they collected data from different sources over
the different surface-brightness ranges.  As they note, they used data
for optically bright galaxies and combined it with LSB objects selected
to complement that data set. This enhances the number of uncataloged LSB
objects, which generally have higher relative gas content, but it may also
miss the occasional magnitude-limited source with a faint disk or HI-selected
source with a relatively bright disk, as are seen in our scatter plot.

We find a much stronger correlation between $L_I/M_{HI}$ and the overall
galaxy surface brightness than with the disk component
alone. McGaugh (1996) has pointed out the problems with defining a surface
brightness using total magnitudes and angular areas, the basic problem
being that the isophotal radius measures the light over an essentially
arbitrary portion of a galaxy's disk. However, a robust measure of the
average surface brightness can be made within the half-light or
effective radius:
   $$\mu_{1/2}=I_{tot}+2.5\log(2\pi r_{1/2}^2)\ .$$
This relationship, shown in Fig.~\ref{sbcorr}(b), has $\rho=0.82$, which
is stronger than any of the correlations found by McGaugh \& de Blok (1997).

Because our $I$--band data did not have very good seeing, we
have used the $B$--band effective radii, but this should make little
difference. The correlation is only slightly diminished if we use
$B$-band values instead ($\rho=0.76$), or use mixes of $B$-- and $I$--band data
for the surface brightness or star-to-gas ratio ($\rho=0.81$ and 0.70).
The correlation is noticeably weaker for the magnitude-selected set of
galaxies alone, although still significant: $\rho=0.62$ using $I$ values;
$\rho=0.54$ using $B$.  This reduced correlation is like the effect McGaugh \&
de Blok found when they excluded LSB galaxies from their sample,
but the values we find for our magnitude-limited sample is still comparable
to the strongest correlations they found for their full sample. We
therefore believe this measure of surface brightness is a more powerful
predictor of the relative gas and stellar content of galaxies.

\subsection{Color Correlations}

Another correlation that has been noted previously (Schneider 1996;
McGaugh \& de Blok 1997) is between the galaxy color and the star-to-gas
ratio. Again, this is in the sense one might expect, that
galaxies with apparently more conversion of gas into stars appear
redder. The correlation we find is stronger than that noted by 
McGaugh \& de Blok, although not quite as strong as the correlation with
surface brightness.
The data are shown in Fig.~\ref{sbcorr}(c), and correlation coefficient
between the $(B-R)$ color and $\log(L_I/M_{HI})$ is $\rho=0.72$
($\rho=0.71$ for the magnitude-selected objects alone).

The correlation substantially weakens when using the blue luminosity
in the star-to-gas ratio. It drops to $\rho=0.51$. We suspect that this
difference primarily reflects that $I$-band is the better measure of the
galaxies' stellar content.
These correlations 
all appear to be consistent with a paradigm in which the galaxies
evolve as gas is consumed to make stars. This is clearly an
oversimplification since the objects we are examining span a wide
range of masses and sizes, with the smallest objects showing the
least gas consumption. However this paradigm may be appropriate
for a scenario in which larger galaxies form from the accretion
of smaller galaxies.

Turning back to Fig.~\ref{briplot}, note the color evolution curves
marked on the plot. These are based on the population synthesis code of
Bruzual \& Charlot (1993, 1998) for a population of stars that has a
single burst of star formation.\footnote{We used the Salpeter IMF over
the full mass range, but the form of the initial mass function used
appeared to make little difference here.
We found that all of their models lie slightly above the
observed locus of points at the blue end of the distribution.}
The log of the age of the population in years after the burst is marked
along each curve, which show results for two different metallicities:
solar (black line) and 1/50th solar (gray line).

Both curves closely follow the empirical relationship between $(B-R)$
and $(B-I)$ found earlier.  Since the evolutionary
tracks are nearly linear in the color-color diagram, any pattern
of star-forming histories should also lie along this line.
The bluest objects have colors comparable to a starburst $<10^9$ yr ago.
Combinations of such a starburst with a pre-existing population would
yield redder colors, suggesting an even younger starburst, but it should
be kept in mind that a starburst only $10^8$ yr old can generate many
times more light per unit mass than a population $>10^9$ yr old.
Thus we do not suggest using the colors as a simple age indicator so
much as a relative indication of the amount of star formation that
has taken place recently.

\section{Summary}

We have presented data from a large, sensitive search for extragalactic
sources in the 21 cm line of neutral hydrogen, which has detected some of
the lowest mass HI sources found in a blind survey to date. The survey
was made with the Arecibo radio telescope and covers about 55 deg$^2$
out to over 100 Mpc. The survey region is in the general vicinity of the
Pisces-Perseus supercluster, crossing through the supergalactic plane, and
it is all at high enough Galactic latitudes ($|b|>25^\circ$) that the
extinction is moderate ($A_B\lesssim1$).
We have obtained $BRI$ photometry for nearly the entire sample, and examined
other objects found within the same volume. In total, 75 objects
were found by the HI survey, 39\% of these being previously unidentified
objects.

Cross-references to other catalogs allow us to examine selection criteria
of the different survey techniques and how they relate to physical
characteristics of the galaxies. In general, we find that the optical
survey techniques tend to collect strongly nucleated, high surface
brightness galaxies. The newly detected HI sources appear to represent
a population of faint blue dwarfs, often with substantially
larger gas than stellar masses.

There are several interesting aspects of the HI-selected sample:

(1) At low redshifts, where the HI survey is sensitive to low mass objects,
we detect three new objects at $cz<1000$km s$^{-1}$, one of which (\#75)
has no clear optical counterpart (although it is too near a bright
star to rule out a very low surface brightness counterpart).
Out to $cz<3000$km s$^{-1}$, 6 of 9 sources are new detections.

(2) The new detections in this survey are, not-surprisingly, HI-bright
relative to the objects found in previous optical surveys. They are also 
gas-dominated objects based on estimations of their total stellar mass
from their $I$-band photometry. Thirteen objects, spanning a wide range
of masses, have a star-to-gas mass ratio less than 1/3 not including
source \#75 whose optical counterpart has not been identified.
Only 3 of these objects were found previously in
diameter-limited searches for LSB sources.

(3) The distribution of light in the new detections is systematically
different from the optically-selected galaxies. They are much less likely
to have any bulge component, and this tendency is not just a result
of selecting lower-mass objects (\S4.2).

(4) The galaxies show a wide range of colors, but follow color-color
tracks of population synthesis models of stellar evolution closely.
The new
HI-detected sources tend to be bluer on average, and our bluest ten
objects lie at younger ages along the evolution track than any of the
objects drawn from either the complete diameter-limited sample of galaxies of
de Jong \& van der Kruit (1994) or the HI-selected sample of Szomoru
et al. (1994).

(5) We find strong correlations between the galaxy star-to-gas mass ratio and
the mean surface brightness and color of these galaxies. The strongest
correlation is found between the star-to-gas mass ratio and the
mean surface brightness within the half-light radius.
VLA observations of our lowest surface brightness source (\#14) indicate
that its surface density of HI is well below the threshold for star formation.

\begin{acknowledgments}

We would like to thank J.~Rosenberg, F.~Briggs, and M.~Zwaan for many
helpful discussions. We also thank the staffs at Arecibo, Kitt Peak, and
the VLA for their assistance with the observations.
This work was supported in part by NSF Presidential Young Investigator
award AST-9158096.

\end{acknowledgments}

\clearpage
 
\begingroup
\oddsidemargin -.2in
\evensidemargin -.2in

\begin{deluxetable}{@{}l@{ }c@{ }c@{ }c@{ }c@{ }c@{ }c@{ }c@{ }c@{ }c@{ }c@{ }c@{ }}
\renewcommand{\arraystretch}{0.82}
\tablecolumns{12}
\footnotesize
\tablecaption{HI Observations in Arecibo Slice. \label{tbl-1}}
\tablewidth{7.0in}
\tablehead{& & &
{$cz_{hel}$} &
{$w_{50}$}  &
{$w_{20}$} &
 &
{$\int S_{cen}\,dv$} &
{$\int S_{tot}\,dv$} &
{$\int S_{det}\,dv$}  &
{$w_{det}$} &
{offset} \\
{No.} &
\multicolumn{2}{c}{$\alpha$\hfil (1950)\hfil $\delta$}  &
{(km s$^{-1}$)} &
{(km s$^{-1}$)} &
{(km s$^{-1}$)} &
{$N_h$} &
{(Jy km s$^{-1}$)} &
{(Jy km s$^{-1}$)} &
{(Jy km s$^{-1}$)} &
{(km s$^{-1}$)} &
{(arcmin)} \\
 (1)&\multicolumn{2}{c}{(2)}&(3)&(4)&(5)&(6)&(7)&(8)&(9)&(10)&(11)
}
\startdata
\phn1&22~02~07.9&+23~04~13&6206&\phn81&150&3&1.36&1.66&0.80&116&1.7\nl
\phn2&22~25~44.5&+23~07~23&1278&\phn62&\phn82&1&2.66&3.53&1.73&\phn76&0.8\nl
\phn3&22~38~58.2&+23~07~07&7154&\phn87&130&1&4.14&4.78&3.21&142&0.5\nl
\phn4&22~56~31.2&+23~44~15&7189&\phn39&125&3&0.83&1.19&0.54&161&1.7\nl
\phn5&23~01~13.7&+23~25~07&1155&\phn96&111&2&2.75&2.85&0.58&106&1.7\nl
\tablevspace{.5ex}
\phn6&23~05~37.3&+22~59~16&6336&\phn31&\phn50&1&1.46&1.62&1.20&\phn45&0.8\nl
\phn7&23~06~06.9&+23~21~49&4270&130&149&2&2.76&3.38&1.90&139&1.6\nl
\phn8&23~18~22.0&+23~32~22&5961&578&612&2&2.38&3.31&1.96&514&1.3\nl
\phn9&23~20~10.2&+23~06~03&5042&118&174&3&1.38&2.27&0.72&131&1.6\nl
10&23~21~47.3&+23~33~52&5895&122&154&2&1.93&2.10&0.76&140&2.1\nl
\tablevspace{.5ex}
11&23~22~04.2&+23~42~49&5063&264&283&2&2.39&4.15&1.50&274&1.8\nl
12&23~25~12.2&+23~18~51&3403&\phn95&166&1&6.78&8.86&1.95&200&2.2\nl
13&23~25~35.7&+23~15~29&3556&269&294&2&9.43&13.80\phn&4.30&402&1.7\nl
14&23~29~06.3&+23~45~42&4009&\phn59&\phn88&1&0.76&0.96&0.65&132&2.0\nl
15&23~29~31.2&+23~17~01&6989&239&249&2&2.27&2.92&1.47&267&0.9\nl
\tablevspace{.5ex}
16&23~29~58.8&+23~39~35&5095&408&423&2&2.55&3.44&1.66&431&1.6\nl
17&23~31~57.3&+23~38~48&5603&164&189&2&0.92&1.68&0.87&166&0.1\nl
18&23~32~15.2&+23~41~44&5558&111&135&1&2.41&2.69&0.87&138&2.1\nl
19&23~33~21.0&+23~20~55&3054&188&208&2&4.06&5.46&3.14&209&0.8\nl
20&\llap{$\sim$}23~40~43.5&+23~13~51&5355&331&344&2&2.46&3.01&2.37&344&1.2\nl
\tablevspace{.5ex}
21&23~40~46.6&+23~14~08&5271&\phn43&\phn63&1&0.51&0.66&0.55&\phn63&0.4\nl
22&\llap{$\sim$}23~59~05.0&+23~12~46&\llap{$\sim$}4325&\llap{$\sim$}590&---&2&\llap{$\sim$}7.14&\llap{$\sim$}9.45&12.36\phn&623&1.6\nl
23&\llap{$\sim$}23~59~05.0&+23~12~46&\llap{$\sim$}4430&\llap{$\sim$}540&---&2&\llap{$\sim$}4.76&\llap{$\sim$}6.30&12.41\phn&623&1.6\nl
24&00~02~32.8&+23~41~37&6573&133&178&2&2.34&2.72&1.28&196&2.1\nl
25&00~03~49.1&+23~30~59&4574&197&259&2&3.52&4.26&1.76&221&1.3\nl
\tablevspace{.5ex}
26&00~06~20.1&+23~32~13&4528&152&192&3&3.83&4.12&1.78&235&1.6\nl
27&00~11~23.6&+23~41~37&6201&138&154&2&1.29&1.38&1.14&161&0.7\nl
28&00~16~24.3&+23~11~49&4459&221&234&2&5.02&5.59&3.28&243&1.2\nl
29&00~19~27.7&+23~27~32&4479&237&253&2&4.03&4.80&3.31&258&1.0\nl
30&00~23~22.8&+23~39~09&5679&190&198&2&1.10&1.08&0.60&141&1.8\nl
\tablevspace{.5ex}
31&00~24~51.0&+23~26~53&5884&188&259&2&0.93&1.23&0.43&123&1.6\nl
32&00~29~54.7&+23~06~57&4658&213&230&2&3.58&4.56&3.31&249&1.3\nl
33&00~30~59.3&+23~07~21&5190&268&280&2&2.70&2.64&2.12&271&1.2\nl
34&00~33~19.5&+23~45~35&5614&222&238&2&1.33&1.60&0.81&101&1.7\nl
35&00~33~25.1&+23~40~40&5252&524&544&2&3.15&3.08&1.03&525&1.9\nl
\tablevspace{.5ex}
36&00~34~15.7&+23~43~01&4618&549&625&2&6.22&8.30&5.32&657&1.1\nl
37&\llap{$\sim$}00~40~15.5&+23~13~08&7274&342&368&2&2.00&2.10&1.66&174&1.5\nl
38&\llap{$\sim$}00~40~15.5&+23~13~08&7148&114&166&3&1.15&1.15&0.93&\phn98&1.5\nl
39&00~42~12.8&+23~32~09&6650&\phn61&\phn74&3&0.60&0.59&0.53&107&0.9\nl
40&\llap{$\sim$}00~46~54.4&+23~47~37&7157&138&177&2&0.81&1.39&0.63&141&1.9\nl
\tablevspace{.5ex}
41&00~49~15.1&+23~12~34&2622&\phn66&\phn80&2&0.71&0.91&0.54&120&1.9\nl
42&00~53~36.6&+23~46~55&4962&\phn47&\phn55&1&0.44&0.44&0.32&\phn53&1.3\nl
43&00~54~39.2&+23~36~50&4834&171&203&3&3.67&4.41&1.95&213&1.7\nl
44&01~17~15.9&+23~42~06&6777&224&246&2&3.18&4.01&1.52&408&1.9\nl
45&01~28~36.6&+23~41~45&3415&\phn63&\phn98&1&4.69&6.66&2.68&\phn90&1.6\nl
\tablevspace{.5ex}
46&01~32~24.1&+23~45~22&3316&115&144&2&5.03&7.57&2.10&131&2.1\nl
47&01~33~51.3&+23~33~56&\phn563&\phn51&\phn66&1&1.16&1.76&0.56&\phn91&1.6\nl
48&01~42~21.3&+23~31~15&3874&\phn44&\phn58&1&0.54&0.86&0.47&\phn57&1.4\nl
49&01~47~54.7&+23~09~17&5614&332&411&2&1.55&2.09&0.72&165&2.3\nl
50&01~51~05.7&+23~39~58&2914&118&146&2&1.69&2.03&1.11&165&1.3\nl
\tablevspace{.5ex}
51&01~57~07.1&+23~24~23&4919&124&158&3&4.93&5.46&2.98&180&0.6\nl
52&01~57~57.0&+23~30~46&5029&121&139&2&1.41&1.79&0.75&111&1.8\nl
53&01~58~14.8&+23~10~39&5041&171&210&3&2.04&2.49&1.38&180&1.4\nl
54&02~00~32.3&+23~31~05&2832&122&143&2&1.74&2.27&1.25&118&1.4\nl
55&02~00~55.8&+23~47~08&4812&252&294&2&1.79&2.42&0.30&\phn65&2.1\nl
\tablevspace{.5ex}
56&02~06~17.3&+23~36~19&5105&\phn64&\phn82&1&1.26&1.54&0.96&\phn89&1.0\nl
57&02~06~19.8&+23~00~58&8162&169&251&3&2.59&3.11&1.37&160&0.7\nl
58&02~10~47.7&+23~00~49&2452&\phn54&\phn72&1&0.34&0.39&0.19&\phn39&0.5\nl
59&\llap{$\sim$}02~22~20.6&+23~37~08&6574&\phn69&104&1&1.53&2.03&0.86&206&1.7\nl
60&\llap{$\sim$}02~22~31.5&+23~36~15&6631&180&213&2&1.78&2.06&1.40&203&1.0\nl
\tablevspace{.5ex}
61&02~25~29.5&+22~59~26&6399&398&436&2&4.19&5.88&3.28&429&0.9\nl
62&02~25~48.9&+23~34~09&6287&296&315&2&2.30&2.91&1.63&156&1.2\nl
63&02~26~30.7&+23~44~58&5484&247&267&2&1.49&1.85&1.24&265&1.0\nl
64&02~29~34.0&+23~25~35&4159&120&142&2&1.40&1.37&1.27&148&0.8\nl
65&02~30~02.8&+23~06~02&5563&250&266&2&5.31&5.83&4.44&277&1.5\nl
\tablevspace{.5ex}
66&02~30~39.6&+23~15~42&4414&160&176&2&1.40&1.89&1.08&174&1.6\nl
67&02~31~52.7&+23~11~43&4360&606&633&2&9.18&12.48\phn&6.44&633&1.3\nl
68&02~33~18.1&+23~40~54&5649&246&278&2&7.47&9.29&4.93&265&1.5\nl
69&02~34~37.0&+23~05~13&8239&319&368&2&2.70&2.85&0.61&273&1.4\nl
70&02~38~25.3&+23~03~07&7661&174&189&3&0.91&1.34&0.75&199&1.9\nl
\tablevspace{.5ex}
71&02~44~03.7&+23~23~16&6196&345&386&2&2.68&3.67&1.01&417&2.0\nl
72&02~45~00.9&+23~11~56&6054&319&332&2&4.07&4.32&1.82&465&2.2\nl
73&02~45~46.7&+23~03~45&\phn469&\phn30&\phn53&1&0.97&1.07&0.31&\phn34&0.4\nl
74&02~51~49.9&+23~10~41&6071&116&140&2&1.63&1.68&1.16&140&1.1\nl
75&02~54~40.1&+23~35~13&\phn411&\phn31&\phn51&1&0.49&0.64&0.23&\phn61&1.0\nl
\tablevspace{.5ex}
\hline
\tablevspace{.5ex}
\multicolumn{12}{c}{Non-HI-Selected Sources}\nl
\tablevspace{.5ex}
\hline
\tablevspace{.5ex}
76&22~38~50.3&+23~08~18&7148&---&---&---&$confused$&---&---&---&1.9\nl
77&23~00~42.9&+23~29~22&7775&156&161&2&0.47&---&\llap{$\sim$}0.14&\llap{$\sim$}161&2.1\nl
78&23~11~20.5&+23~32~53&6122&108&128&3&0.17&---&\llap{$\sim$}0.07&\llap{$\sim$}128&1.9\nl
79&23~12~01.3&+23~24~47&6048&---&---&---&\llap{$<$}0.10&---&---&---&1.8\nl
80&23~12~24.1&+23~00~39&7935&438&458&2&0.69&---&\llap{$\sim$}0.32&\llap{$\sim$}458&1.5\nl
\tablevspace{.5ex}
81&23~51~06.8&+23~21~16&8184&---&---&---&\llap{$<$}0.18&---&---&---&1.6\nl
82&00~17~14.3&+23~29~44&5829&389&405&2&0.65&---&0.58&356&1.8\nl
83&00~23~20.3&+23~30~21&7389&206&228&2&1.61&2.34&0.60&242&1.9\nl
84&00~24~15.4&+23~14~37&6896&\phn90&106&---&\llap{$\sim$}0.46&---&0.36&106&0.9\nl
85&00~34~13.9&+23~42~34&4595&---&---&---&---&---&---&---&0.7\nl
\tablevspace{.5ex}
86&00~35~53.5&+23~20~13&7566&134&171&2&$confused$&---&\llap{$\sim$}0.06&\llap{$\sim$}171&1.7\nl
87&00~40~24.0&+23~14~00&7571&359&390&2&1.10&1.56&0.80&390&1.4\nl
88&00~41~41.3&+23~33~36&6493&\phn\llap{$\sim$}37&\phn\llap{$\sim$}69&---&\llap{$\sim$}0.75&---&0.44&\phn69&1.4\nl
89&00~56~20.1&+23~34~58&5086&283&300&2&0.52&---&0.28&117&0.9\nl
90&00~56~42.6&+23~42~35&5043&136&173&2&0.38&---&0.47&106&0.6\nl
\tablevspace{.5ex}
91&02~33~13.2&+23~36~40&5734&177&217&2&0.86&1.53&0.42&269&1.7\nl
\enddata
\tablenotetext{}{
\baselineskip14pt \vskip-20pt
\hskip1em Notes.\ --- 
Column (3): mean velocity of 50\%-of-peak points in profile.
Columns (4) and (5): 50\% and 20\%-of-peak widths measured as fraction
of peak at each edge of profile in 2-horn profiles.
Columns (7)--(9): Integrated flux under HI profile for: single-beam
centered on best-fit position; integrated over 7-beam hexagonal map area;
and in original search spectrum.
Column (10): 20\% width as determined in original search-spectrum detection.
Column (11): Offset of detection position from best-fit HI position.
}
 
\end{deluxetable}
 
\clearpage
\endgroup

\clearpage

\begingroup
\oddsidemargin 0in
\evensidemargin 0in
 
\begin{deluxetable}{@{}l@{ }c@{ }c@{ }c@{ }c@{ }c@{ }c}
\renewcommand{\arraystretch}{0.82}
\tablecolumns{5}
\tablewidth3.6in
\footnotesize
\tablecaption{VLA HI Observations. \label{tbl-1a}}
\tablehead{
\vbox to12pt{}&&&offset&&dimensions\\
{No.} &
\multicolumn{2}{c}{$\alpha$\hfil (1950)\hfil $\delta$}  &
(arcmin)&
\vbox to8pt{\vss \hbox{${\int\displaystyle S_{V\!L\!A}\,dv\atop\ }\over{\ \atop\int\displaystyle S_{tot}\,dv}$}\vskip-4pt} & 
(arcmin)&\\
 (1)&\multicolumn{2}{c}{(2)}&(3)&(4)&(5)
} 
\startdata
14&23~29~08.6&+23~45~27&0.58&1.15&$1.5\times1.0$\nl
17&23~31~57.3&+23~37~54&0.90&0.36&$1.1\times0.3$\nl
18&23~32~14.9&+23~41~37&0.14&0.91&$1.1\times0.5$\nl
75&02~54~42.5&+23~35~11&0.55&0.63&$0.0\times0.0$\nl
\enddata
\tablenotetext{}{
\baselineskip14pt \vskip-20pt
\hskip1em Notes.\ --- 
Column (3): offset of VLA position from best-fit Arecibo position.
Column (4): ratio of VLA and Arecibo total fluxes.
Column (5): estimated HI dimensions based on Gaussian fit and deconvolution
from beam.
}
\end{deluxetable}

\clearpage
\endgroup

\begingroup
\clearpage

\topmargin -.5in
\oddsidemargin -.75in
\evensidemargin -.75in
 
\begin{deluxetable}{@{}l@{ }c@{ }c@{ }c@{ }l@{ }l@{}l@{}r@{/}l@{\quad }l@{ }}
\renewcommand{\arraystretch}{0.82}
\tablecolumns{10}
\footnotesize
\tablecaption{Optical Counterparts to the HI-Selected Sample. \label{tbl-2}}
\tablewidth{7.9in}
\tablehead{
& & & &\phm{MCG+00--00--000}&\phm{LSBC~F480--V03}&\phm{Mrk~325/UCM/KUG}&\multicolumn{1}{r}{\phm{0.000/}}&\phm{$<$0.000}&\phm{III~Zw~125/VV~254}\\ 
& & & & \multicolumn{6}{c}{Optical Catalog Cross-Identifications}\\
\cline{5-10}\\
& & &
offset &
Magnitude & 
Diameter & 
UV Excess/ & 
\multicolumn{2}{c}{IRAS FIR}& 
 \\
{No.} &
\multicolumn{2}{c}{$\alpha$\hfil (1950)\hfil $\delta$}  &
{(arcmin)} & 
Limited & 
Limited & 
Emission Line & 
$60\mu$&$100\mu$ & 
Morphology 
\\
(1)&\multicolumn{2}{c}{(2)}&(3)&\quad(4)&\quad(5)&\quad(6)&\multicolumn{2}{c}{(7)}&\quad{(8)}
} 
\startdata
\phn1&22~02~09.1&+23~04~38&0.5&---&---&---&---&---&---\nl
\phn2&22~25~44.6&+23~07~07&0.3&---&LSBC~F533--01&---&---&---&Im\nl
\phn3&22~39~00.3&+23~06~50&0.6&IC~5243&UGC~12153&Ark~562/KUG&0.538&$<$2.538&S? pec\nl
\phn4&22~56~33.6&+23~44~09&0.5&---&---&---&---&---&---\nl
\phn5&23~01~14.8&+23~24~59&0.3&---&ESDO~535--04&---&---&---&---\nl
\tablevspace{.5ex}
\phn6&23~05~38.7&+22~59~24&0.3&---&---&---&---&---&---\nl
\phn7&23~06~05.8&+23~21~38&0.3&---&---&---&---&---&---\nl
\phn8&23~18~24.1&+23~32~00&0.6&CGCG~476--013&UGC~12533&---&---&---&Sbc\nl
\phn9&23~20~08.0&+23~06~02&0.5&CGCG~476--023&---&---&---&---&S\nl
10&23~21~47.7&+23~34~08&0.3&---&---&---&---&---&---\nl
\tablevspace{.5ex}
11&23~22~04.9&+23~42~42&0.2&CGCG~476--031&UGC~12583&---&---&---&Scd\nl
12&23~25~11.7&+23~18~49&0.1&NGC~7673&UGC~12607&Mrk~325/UCM/KUG&4.913&6.893&(R$'$)SAc? pec\nl
13&23~25~36.6&+23~15~22&0.2&NGC~7677&UGC~12610&Mrk~326/KUG&3.956&5.915&SAB(rc)bc: pec\nl
14&23~29~08.9&+23~45~38&0.6&---&---&---&---&---&---\nl
15&23~29~32.1&+23~16~55&0.2&CGCG~476--061&---&KUG&0.301&$<$1.111&S?\nl
\tablevspace{.5ex}
16&23~29~59.7&+23~39~48&0.3&CGCG~476--064&UGC~12655&---&---&---&S0\nl
17&23~31~56.9&+23~37~39&1.2&---&---&---&---&---&---\nl
18&23~32~14.3&+23~41~42&0.2&---&---&---&---&---&---\nl
19&23~33~21.0&+23~20~32&0.4&NGC~7712&UGC~12694&KUG&0.656&1.860&E?\nl
20&23~40~43.0&+23~14~02&\llap{$\sim$}0.2&---&FGC~2530&---&---&---&Scd\nl
\tablevspace{.5ex}
21&23~40~46.0&+23~13~19&\llap{$\sim$}0.8&---&---&---&---&---&---\nl
22&23~59~04.4&+23~12~22&\llap{$\sim$}0.4&CGCG~477--040&UGC~12914&Kaz~240&6.128&14.010&(R)S(r)cd: pec\nl
23&23~59~08.0&+23~13~07&\llap{$\sim$}0.8&CGCG~477--041&UGC~12915&---&6.128&14.010&S? pec\nl
24&00~02~30.7&+23~41~30&0.5&---&---&---&---&---&---\nl
25&00~03~48.4&+23~30~38&0.4&---&---&KUG&---&---&S\nl
\tablevspace{.5ex}
26&00~06~19.7&+23~32~21&0.2&NGC~9&UGC~78&KUG/UCM&0.737&1.527&Sb: pec\nl
27&00~11~23.5&+23~41~28&0.2&---&---&KUG&---&---&S\nl
28&00~16~24.4&+23~12~02&0.2&CGCG~478--053&UGC~179&---&---&---&Scd:\nl
29&00~19~28.6&+23~27~33&0.2&CGCG~479--016&UGC~210&---&---&---&Sb\nl
30&00~23~20.0&+23~38~49&0.7&---&LSBC~F473--01&---&---&---&Im\nl
\tablevspace{.5ex}
31&00~24~50.8&+23~26~46&0.1&---&---&---&---&---&---\nl
32&00~29~54.7&+23~07~10&0.2&CGCG~479--037&UGC~321&---&---&---&SBcd?\nl
33&00~30~58.1&+23~07~20&0.3&CGCG~479--040&---&---&---&---&S?\nl
34&00~33~19.7&+23~45~43&0.1&MCG+04--02--032&UGC~354&---&---&---&---\nl
35&00~33~25.7&+23~40~58&0.3&NGC~160&UGC~356&---&---&---&(R)SA0+ pec\nl
\tablevspace{.5ex}
36&00~34~13.5&+23~42~58&0.5&NGC~169&UGC~365&---&1.118&3.668&SA(s)ab: pec\nl
37&00~40~16.7&+23~12~50&\llap{$\sim$}0.4&CGCG~479--061&---&UCM&0.783&2.233&SB\nl
38&00~40~18.7&+23~13~29&\llap{$\sim$}0.8&NGC~228&UGC~458&---&---&---&(R)SB(r)ab\nl
39&00~42~12.6&+23~32~00&0.2&---&---&---&---&---&---\nl
40&00~47~02.8&+23~46~24&\llap{$\sim$}2.3&---&---&---&---&---&---\nl
\tablevspace{.5ex}
41&00~49~14.6&+23~12~35&0.1&---&---&---&---&---&---\nl
42&00~53~35.7&+23~46~59&0.2&---&---&---&---&---&---\nl
43&00~54~38.7&+23~37~10&0.3&CGCG~480--025&UGC~591&UCM&---&---&S?\nl
44&01~17~15.2&+23~42~09&0.2&MCG+04--04--003&---&---&---&---&---\nl
45&01~28~37.0&+23~41~48&0.1&---&UGC~1084&---&---&---&Sm:\nl
\tablevspace{.5ex}
46&01~32~24.6&+23~45~06&0.3&---&---&---&---&---&---\nl
47&01~33~50.3&+23~33~37&0.4&---&---&---&---&---&---\nl
48&01~42~21.5&+23~31~08&0.1&---&---&---&---&---&---\nl
49&01~47~55.7&+23~09~08&0.3&---&---&UCM&---&---&Sa\nl
50&01~51~06.6&+23~40~02&0.2&---&---&---&---&---&---\nl
\tablevspace{.5ex}
51&01~57~06.1&+23~24~07&0.3&NGC~776&UGC~1471&UCM&1.250&3.099&SAB(rs)b\nl
52&01~57~56.1&+23~30~56&0.3&---&---&---&---&---&---\nl
53&01~58~15.2&+23~10~37&0.1&CGCG~482--046&---&---&0.248&0.818&S\nl
54&02~00~32.4&+23~31~16&0.2&---&UGC~1538&---&---&---&Im:\nl
55&02~00~54.3&+23~47~18&0.4&---&---&---&---&---&---\nl
\tablevspace{.5ex}
56&02~06~17.8&+23~36~23&0.1&---&---&---&---&---&---\nl
57&02~06~22.6&+23~00~56&0.7&---&---&UCM&0.602&$<$1.369&Sc+\nl
58&02~10~46.7&+23~01~29&0.7&---&---&---&---&---&---\nl
59&02~22~19.1&+23~37~36&\llap{$\sim$}0.6&CGCG~483--034&---&---&---&---&S?\nl
60&02~22~31.0&+23~36~08&\llap{$\sim$}0.2&CGCG~483--036&---&---&0.335&$<$2.598&---\nl
\tablevspace{.5ex}
61&02~25~31.3&+22~59~29&0.4&CGCG~483--063&UGC~1938&---&0.293&$<$1.688&Sbc\nl
62&02~25~48.5&+23~34~24&0.3&CGCG~483--065&UGC~1950&---&0.360&$<$1.631&S?\nl
63&02~26~29.7&+23~44~54&0.2&---&---&---&---&---&---\nl
64&02~29~34.4&+23~25~58&0.4&---&---&---&---&---&---\nl
65&02~30~02.8&+23~06~27&0.4&CGCG~484--005&UGC~2020&---&---&---&Scd:\nl
\tablevspace{.5ex}
66&02~30~37.9&+23~15~57&0.5&---&---&---&---&---&---\nl
67&02~31~51.5&+23~11~41&0.3&NGC~984&UGC~2059&---&---&---&SA0+ pec\nl
68&02~33~17.5&+23~40~55&0.1&CGCG~484--012&UGC~2079&---&0.513&1.522&SAB(s)c\nl
69&02~34~35.5&+23~05~02&0.4&CGCG~484--016&UGC~2104&---&0.386&$<$1.408&Scd:\nl
70&02~38~26.7&+23~03~18&0.4&---&---&---&---&---&---\nl
\tablevspace{.5ex}
71&02~44~01.8&+23~23~10&0.4&CGCG~484--018&UGC~2248&---&0.847&1.955&S0:\nl
72&02~44~58.8&+23~11~49&0.5&CGCG~484--019&UGC~2267&---&---&---&(R)SB(r)b\nl
73&02~45~46.3&+23~03~58&0.2&---&---&---&---&---&---\nl
74&02~51~49.0&+23~10~28&0.3&---&LSBC~F480-V03&---&---&---&SBm\nl
75&\multicolumn{2}{c}{---}&---&---&---&---&---&---&---\nl
\tablevspace{.5ex}
\hline
\tablevspace{.5ex}
\multicolumn{9}{c}{Non-HI-Selected Sources}\nl
\tablevspace{.5ex}
\hline
\tablevspace{.5ex}
76&22~38~51.3&+23~08~47&---&IC~5242&UGC~12148&UCM&1.109&$<$3.121&Sa\nl
77&23~00~42.9&+23~29~22&---&CGCG~475--028&---&---&3.641&5.515&---\nl
78&23~11~20.5&+23~32~53&---&---&---&Mrk~317/KUG&0.667&$<$2.799&S?\nl
79&23~12~01.4&+23~24~45&---&NGC~7539&UGC~12443&---&---&---&S0\nl
80&23~12~25.0&+23~00~37&---&---&UGC~12449&---&---&---&---\nl
\tablevspace{.5ex}
81&23~51~06.8&+23~21~16&---&---&---&UCM&---&---&---\nl
82&00~17~12.7&+23~29~42&---&IC~1540&UGC~186&---&---&---&SBb\nl
83&00~23~20.2&+23~30~29&0.1&---&---&---&---&---&---\nl
84&00~24~15.4&+23~14~37&---&---&---&HS&---&---&---\nl
85&00~34~13.9&+23~42~34&---&IC~1559&---&Mrk~341&1.118&3.668&SAB pec:\nl
\tablevspace{.5ex}
86&00~35~53.6&+23~20~20&---&CGCG~479--049&---&Mrk~344&---&---&E?\nl
87&00~40~25.7&+23~14~08&---&NGC~229&---&---&---&---&---\nl
88&00~41~41.3&+23~33~36&---&---&---&HS&---&---&---\nl
89&00~56~19.9&+23~34~58&---&CGCG~480--028&UGC~612&---&0.339&1.475&S?\nl
90&00~56~42.5&+23~42~35&---&CGCG~480--029&---&---&---&---&Sc\nl
\tablevspace{.5ex}
91&02~33~12.2&+23~36~48&0.3&---&---&---&0.214&$<$0.942&---\nl
\enddata
\tablenotetext{}{
\baselineskip14pt \vskip-20pt
\hskip1em Notes.\ --- 
Column (3): offset of optical position from best-fit Arecibo HI position.
Column (4): NGC and IC (Dreyer 1888, 1895, 1908); CGCG (Zwicky et al.~1961--8);
MCG (Vorontsov-Velyaminov et al.~1962--8).
Column (5): UGC (Nilson 1973); ESDO (Eder et al.~1989); LSBC
(Schombert et al.~1992); FGC (Karachentsev et al.~1993).
Column (6): Mrk (Markaryan et al. 1989); Ark (Arakelian 1975);
Kaz (Kazaryan 1979); UCM (Zamorano et al.~1994, Rego et al.~1993);
KUG (Takase \& Miyauchi-Isobe 1993); HS (Popescu et al.~1996).
Column (7): IRAS 60 and 100$\mu$ fluxes for detected sources.
Column (8): morphological types from NED.
}

\end{deluxetable}

\clearpage
\endgroup

\begingroup
\clearpage

\topmargin -.5in
\oddsidemargin -.75in
\evensidemargin -.75in
 
\begin{deluxetable}{@{}l@{ }c@{\quad}c@{\quad}c@{\quad}c@{\quad}c@{\quad}c@{\quad}c@{\quad}c@{ }c@{ }c@{ }c@{ }c@{ }c@{\quad}c@{\quad}c@{\quad }c@{\quad}c@{\quad}c}
\renewcommand{\arraystretch}{0.82}
\tablecolumns{18}
\footnotesize
\tablecaption{Optical Measurements of the HI-Selected Sample. \label{tbl-3}}
\tablewidth{7.75in}
\tablehead{
{No.} &
$B_{25}$ &
$A_B$ & 
$B^0_{25}$ & 
$B^0_T$ &
$a_{25}$ & 
$a_H$ & 
$a_{1/2}$ & 
$(b/a)$ &
$(b/a)_H$ &
P.A. &
$(B$$-$$R)$ &
$(B$$-$$I)$ &
$(B$$-$$R)_0$ &
$\mu_0$ &
$f_B$ &
$\mu_d$ &
$r_d$
\\
(1)&(2)&(3)&(4)&(5)&(6)&(7)&(8)&(9)&(10)&(11)&
(12)&(13)&(14)&(15)&(16)&(17)&(18)
} 
\startdata
\phn1&16.90&0.40&16.51&16.42&0.53\rlap{$'$}&0.67\rlap{$'$}&0.30\rlap{$'$}&0.80&0.80&129\rlap{$^\circ$}&0.86&1.16&0.75&22.05&---&22.45&\phn7.6\rlap{$''$}\nl
\phn2&18.07&0.23&18.05&16.94&0.39&0.91&0.58&0.69&0.82&\phn62&0.73&0.86&0.81&23.72&---&23.85&\phn9.8\nl
\phn3&14.66&0.26&14.41&14.39&0.74&0.89&0.30&0.86&0.89&\phn11&0.69&\llap{$\sim$}0.97&0.70&18.88&0.90$^q$&---&---\nl
\phn4&16.97&0.65&16.54&16.29&0.60&0.85&0.34&0.21&0.27&142&1.08&1.33&1.15&21.61&---&22.36&\phn5.9\nl
\phn5&17.54&0.56&17.11&16.65&0.47&0.83&0.38&0.48&0.53&133&0.78&1.17&0.79&22.37&---&22.81&\phn6.5\nl
\tablevspace{.5ex}
\phn6&19.95&0.83&19.17&17.51&0.21&0.69&0.50&0.44&0.92&\phn78&0.73&1.08&0.89&22.85&0.06$^e$&24.36&\phn9.1\nl
\phn7&17.32&0.78&16.48&15.98&0.58&0.99&0.46&0.57&0.63&110&1.01&1.46&1.30&22.13&0.03$^e$&22.91&\phn8.0\nl
\phn8&15.67&0.38&15.67&15.24&0.86&1.67&0.61&0.20&0.17&\phn18&1.47&\llap{$\sim$}2.07&1.75&20.51&0.45$^q$&23.05&11.9\nl
\phn9&16.10&0.27&15.84&15.74&0.52&0.75&0.22&0.84&0.84&179&0.78&1.13&1.00&21.00&---&20.86&\phn3.8\nl
10&18.27&0.20&18.50&17.81&0.27&0.58&0.29&0.57&0.61&124&0.88&1.08&0.92&22.81&---&23.23&\phn5.0\nl
\tablevspace{.5ex}
11&16.18&0.20&16.40&16.04&0.74&1.19&0.51&0.18&0.18&\phn53&---&---&---&21.61&0.03$^e$&22.48&\phn8.6\nl
12&13.35&0.19&13.17&13.10&1.25&1.93&0.26&0.83&0.84&\phn83&0.68&0.99&0.73&18.33&0.77$^e$&22.70&14.8\nl
13&14.18&0.18&14.08&13.96&1.48&2.03&0.50&0.54&0.60&\phn37&1.12&1.62&1.43&19.91&0.40$^e$&22.32&14.1\nl
14&23.10&0.25&25.96&18.47&0.00&0.51&0.48&0.85&0.85&\phn82&0.90&\llap{$\sim$}1.28&0.91&25.02&---&25.06&\phn9.6\nl
15&15.71&0.28&15.44&15.23&0.53&0.95&0.22&0.81&0.68&\phn88&1.10&1.62&1.29&20.05&0.44$^e$&22.33&\phn6.6\nl
\tablevspace{.5ex}
16&14.23&0.24&14.04&13.94&1.13&1.64&0.34&0.59&0.58&125&1.53&2.04&1.71&18.92&0.37$^e$&21.40&\phn9.5\nl
17&17.11&0.32&16.88&16.77&0.38&0.53&0.20&0.43&0.61&\phn49&0.74&1.01&0.78&21.22&---&21.50&\phn3.6\nl
18&18.00&0.32&17.80&17.44&0.33&0.56&0.25&0.62&0.72&133&0.60&0.65&0.68&22.39&---&22.60&\phn4.0\nl
19&14.26&0.23&14.03&13.99&0.97&1.25&0.42&0.81&0.94&104&1.14&1.61&1.61&20.06&0.60$^q$&---&---\nl
20&16.96&0.26&17.45&16.92&0.51&0.91&0.46&0.15&0.22&124&1.28&1.87&1.60&22.21&---&23.29&\phn8.9\nl
\tablevspace{.5ex}
21&18.45&0.26&18.29&17.99&0.25&0.41&0.18&0.52&0.60&\phn73&0.59&0.79&0.10&22.17&---&22.46&\phn2.8\nl
22&13.73&0.48&13.29&13.11&2.09&2.95&0.98&0.59&0.75&159&1.37&1.99&1.82&18.82&0.18$^q$&22.17&20.0\nl
23&14.71&0.48&14.30&14.17&1.36&1.89&0.58&0.27&0.27&134&1.29&1.88&1.02&20.10&0.03$^e$&21.29&10.6\nl
24&17.47&0.56&16.90&16.51&0.53&0.85&0.39&0.87&0.99&138&0.95&1.35&1.29&22.40&---&22.86&\phn6.7\nl
25&16.43&0.64&15.86&15.64&0.68&1.03&0.39&0.41&0.33&\phn38&0.89&1.11&0.85&21.26&0.06$^e$&22.04&\phn6.7\nl
\tablevspace{.5ex}
26&15.01&0.40&14.67&14.54&0.86&1.33&0.34&0.49&0.44&156&0.87&1.18&0.85&20.02&0.64$^e$&22.70&\phn9.9\nl
27&15.89&0.21&15.80&15.67&0.73&1.03&0.34&0.41&0.43&130&0.90&1.26&0.92&21.24&---&21.52&\phn6.0\nl
28&15.45&0.28&15.25&15.05&0.85&1.29&0.42&0.54&0.46&176&0.87&1.24&1.29&20.55&0.09$^q$&21.72&\phn7.8\nl
29&15.35&0.18&15.24&15.13&0.92&1.19&0.46&0.42&0.42&\phn18&1.03&1.28&1.23&20.70&0.26$^e$&22.75&13.2\nl
30&17.99&0.14&18.06&17.43&0.35&0.71&0.39&0.77&0.80&148&0.98&1.28&1.05&23.20&---&23.40&\phn6.5\nl
\tablevspace{.5ex}
31&16.77&0.14&16.70&16.56&0.48&0.67&0.22&0.72&0.72&133&0.89&1.26&0.95&21.53&---&21.56&\phn4.0\nl
32&15.87&0.12&15.89&15.53&0.77&1.25&0.50&0.30&0.24&148&0.93&1.40&1.25&21.42&0.04$^e$&22.42&\phn8.9\nl
33&15.68&0.12&15.65&15.53&0.73&0.97&0.42&0.47&0.45&128&1.09&1.57&1.35&21.35&0.08$^e$&22.30&10.6\nl
34&16.38&0.14&16.54&16.26&0.60&0.94&0.39&0.30&0.30&116&1.07&1.38&1.25&21.68&0.01$^e$&22.35&\phn6.5\nl
35&13.59&0.13&13.55&13.42&1.89&2.57&0.84&0.51&0.55&\phn45&1.45&1.99&1.55&18.83&0.82$^q$&---&---\nl
\tablevspace{.5ex}
36&14.31&0.12&14.44&14.06&1.26&2.44&0.79&0.36&0.41&\phn85&1.79&2.42&2.25&19.88&0.32$^e$&23.05&20.5\nl
37&15.86&0.16&15.86&15.56&0.65&1.05&0.42&0.40&0.45&\phn51&1.54&2.18&1.94&21.00&0.05$^e$&22.17&\phn7.1\nl
38&14.99&0.16&14.86&14.61&0.83&1.51&0.42&0.76&0.93&112&1.46&2.11&1.74&19.13&0.36$^q$&22.27&10.1\nl
39&18.32&0.15&18.57&18.15&0.24&0.44&0.20&0.74&0.87&135&0.66&0.78&0.64&22.61&---&22.68&\phn3.3\nl
40&16.56&0.17&16.50&16.33&0.39&0.67&0.16&0.79&0.80&\phn69&1.58&2.07&1.43&20.49&0.61$^e$&23.36&\phn6.0\nl
\tablevspace{.5ex}
41&16.26&0.14&16.18&16.11&0.42&0.63&0.14&0.55&0.69&\phn93&0.67&1.01&0.65&20.11&0.47$^e$&21.70&\phn3.8\nl
42&17.86&0.23&18.32&17.56&0.32&0.65&0.39&0.49&0.88&153&1.00&1.24&0.91&23.07&---&23.47&\phn6.1\nl
43&15.42&0.20&15.27&15.19&0.75&1.05&0.34&0.61&0.68&155&0.77&1.11&0.72&20.71&---&21.09&\phn5.7\nl
44&16.82&0.30&16.69&16.34&0.47&0.83&0.34&0.85&0.85&\phn99&0.90&1.31&1.09&22.13&---&22.09&\phn5.4\nl
45&17.61&0.41&18.32&16.73&0.34&1.26&0.70&0.46&0.81&175&0.92&\llap{$\sim$}1.30&0.95&23.16&0.17$^e$&24.79&20.7\nl
\tablevspace{.5ex}
46&17.12&0.48&16.61&16.48&0.41&0.60&0.20&0.93&0.94&\phn66&1.05&1.34&1.12&21.54&---&21.31&\phn3.4\nl
47&18.68&0.49&19.04&18.10&0.25&0.65&0.39&0.41&0.47&\phn63&0.55&0.86&0.61&23.38&---&23.83&\phn6.8\nl
48&18.73&0.49&18.20&17.64&0.31&0.58&0.29&0.96&0.95&\phn64&0.74&1.03&0.81&23.30&---&23.40&\phn5.3\nl
49&16.92&0.44&16.53&16.36&0.40&0.65&0.14&0.46&0.47&177&1.02&1.43&0.93&19.87&0.47$^e$&22.55&\phn4.8\nl
50&17.24&0.50&16.91&16.58&0.53&0.87&0.39&0.49&0.47&159&0.85&1.21&1.01&22.36&---&22.51&\phn6.4\nl
\tablevspace{.5ex}
51&13.90&0.41&13.47&13.39&1.50\rlap{$'$}&2.03\rlap{$'$}&0.50\rlap{$'$}&0.93&0.86&\phn50\rlap{$^\circ$}&1.26&1.82&1.64&18.88&0.19$^q$&21.09&10.9\rlap{$''$}\nl
52&17.73&0.39&17.52&17.09&0.43&0.74&0.34&0.59&0.58&\phn98&0.61&1.01&0.75&22.85&---&22.85&\phn5.9\nl
53&15.72&0.41&15.31&15.24&0.53&0.75&0.22&0.69&0.69&\phn\phn4&0.93&1.37&1.04&20.36&---&20.19&\phn3.5\nl
54&16.61&0.40&16.36&16.16&0.68&0.94&0.48&0.61&0.67&\phn59&0.89&1.32&0.95&22.49&0.01$^e$&23.11&11.8\nl
55&16.83&0.36&16.55&16.34&0.54&0.80&0.39&0.68&0.68&\phn70&1.02&1.38&1.26&22.26&---&22.86&\phn9.3\nl
\tablevspace{.5ex}
56&17.87&0.32&18.00&17.48&0.33&0.62&0.29&0.47&0.51&\phn16&0.78&1.16&0.68&22.72&---&22.88&\phn5.0\nl
57&16.25&0.39&15.94&15.86&0.59&0.79&0.26&0.25&0.32&136&0.71&1.05&0.85&20.16&0.02$^e$&20.88&\phn4.1\nl
58&18.47&0.36&18.45&17.72&0.32&0.65&0.34&0.43&0.44&\phn30&0.90&1.43&0.66&23.18&---&23.60&\phn6.4\nl
59&16.16&0.55&15.63&15.42&0.67&1.08&0.39&0.67&0.67&\phn66&0.83&1.21&1.09&21.63&---&21.79&\phn6.6\nl
60&15.71&0.56&15.14&15.00&0.64&1.14&0.25&0.86&0.86&157&0.92&1.40&0.99&20.29&0.78$^e$&24.24&14.4\nl
\tablevspace{.5ex}
61&15.82&0.61&15.25&15.10&0.90&1.25&0.42&0.28&0.33&154&1.41&1.96&1.67&19.90&0.08$^e$&21.56&\phn7.4\nl
62&15.58&0.52&15.15&14.91&0.89&1.48&0.39&0.50&0.45&144&1.18&1.74&1.40&20.52&0.44$^e$&23.10&12.6\nl
63&16.38&0.54&15.77&15.55&0.77&1.10&0.39&0.27&0.28&110&0.93&1.35&1.20&21.15&0.16$^e$&22.50&\phn8.0\nl
64&17.77&0.61&17.34&16.96&0.43&0.73&0.30&0.52&0.59&\phn94&0.75&1.20&0.57&21.85&0.07$^e$&22.83&\phn5.7\nl
65&16.22&0.72&15.46&15.19&0.85&1.37&0.50&0.28&0.26&108&0.81&1.19&1.03&20.90&0.07$^e$&22.31&\phn9.5\nl
\tablevspace{.5ex}
66&17.55&0.69&17.13&16.63&0.47&0.90&0.39&0.37&0.26&112&1.19&1.60&1.14&22.05&---&22.64&\phn6.7\nl
67&14.20&0.66&13.51&13.38&1.66&2.49&0.58&0.65&0.72&117&1.52&2.17&1.85&18.40&0.33$^q$&21.71&15.2\nl
68&14.74&0.58&14.29&14.11&1.58&2.21&0.84&0.46&0.48&153&0.92&1.36&1.18&21.26&0.02$^e$&21.99&14.3\nl
69&15.70&0.72&14.97&14.82&0.95&1.35&0.46&0.59&0.57&\phn52&1.08&1.60&1.73&20.80&0.03$^e$&21.61&\phn8.0\nl
70&18.46&0.78&17.77&17.44&0.34&0.59&0.26&0.46&0.44&\phn10&0.53&0.78&0.80&22.46&---&22.54&\phn4.2\nl
\tablevspace{.5ex}
71&15.83&1.01&14.77&14.65&0.86&1.27&0.34&0.58&0.58&149&1.32&1.91&1.62&19.28&0.31$^q$&21.37&\phn6.9\nl
72&15.42&1.02&14.32&14.12&1.14&2.25&0.50&0.75&0.76&149&1.34&1.95&1.71&19.95&1.00$^q$&---&---\nl
73&19.19&1.01&18.09&17.76&0.32&0.51&0.23&0.65&0.74&121&0.48&0.44&0.42&22.71&---&22.61&\phn3.8\nl
74&19.24&0.92&18.51&17.52&0.27&0.71&0.38&0.38&0.80&165&0.95&1.18&0.85&22.48&---&23.62&\phn7.0\nl
75&---&0.75&---&---&---&---&---&---&---&---&---&---&---&---&---&---&---\nl
\tablevspace{.5ex}
\hline
\tablevspace{.5ex}
\multicolumn{18}{c}{Non-HI-Selected Sources}\nl
\tablevspace{.5ex}
\hline
\tablevspace{.5ex}
76&14.86&0.26&14.67&14.58&0.85&1.19&0.30&0.83&0.82&157&1.10&\llap{$\sim$}1.55&1.20&19.42&0.25$^e$&21.13&\phn6.5\nl
77&17.91&0.57&17.41&17.05&0.40&0.67&0.29&0.66&0.66&\phn60&\llap{$\sim$}2.46&3.46&---&22.00&---&22.71&\phn5.0\nl
78&16.87&0.58&16.30&16.06&0.44&0.78&0.16&0.60&0.74&110&1.92&2.62&1.78&19.75&1.00$^q$&---&---\nl
79&14.07&0.45&13.62&13.49&1.25&2.05&0.34&0.84&0.89&139&1.42&2.01&1.63&18.17&0.87$^q$&21.96&\phn6.8\nl
80&16.37&0.47&16.09&15.96&0.66&0.91&0.30&0.20&0.22&140&1.88&\llap{$\sim$}2.65&1.92&20.51&0.07$^e$&21.59&\phn5.4\nl
\tablevspace{.5ex}
81&---&0.42&---&---&---&---&---&---&---&---&---&---&---&---&---&---&---\nl
82&15.09&0.20&14.94&14.85&0.91&1.21&0.34&0.43&0.42&\phn27&1.48&1.97&2.62&20.18&0.11$^e$&21.21&\phn6.7\nl
83&---&0.17&---&---&---&---&---&---&---&---&---&---&---&---&---&---&---\nl
84&---&0.19&---&---&---&---&---&---&---&---&---&---&---&---&---&---&---\nl
85&14.82&0.12&14.76&14.39&0.79&1.78&0.39&0.72&0.69&163&1.28&1.83&1.25&19.35&1.00$^q$&---&---\nl
\tablevspace{.5ex}
86&16.67&0.11&16.64&16.32&0.40&0.80&0.20&0.80&0.85&\phn64&1.47&2.18&1.56&20.55&1.00$^q$&---&---\nl
87&14.78&0.16&14.73&14.58&0.76&1.33&0.18&0.36&0.77&\phn99&1.34&1.80&0.94&18.16&1.00$^q$&---&---\nl
88&---&0.15&---&---&---&---&---&---&---&---&---&---&---&---&---&---&---\nl
89&---&0.19&---&---&---&---&---&---&---&---&---&---&---&---&---&---&---\nl
90&---&0.18&---&---&---&---&---&---&---&---&---&---&---&---&---&---&---\nl
\tablevspace{.5ex}
91&16.80&0.58&16.38&16.19&0.62&0.92&0.34&0.28&0.28&100&0.92&1.24&0.78&20.96&0.78$^q$&---&---\nl
\enddata
\tablenotetext{}{
\baselineskip14pt \vskip-20pt
\hskip1em Notes.\ --- 
Entries in columns (4)--(18) are corrected for extinction and
inclination: isophotes are adjusted as described in the text.
Column (5): total magnitude based on extrapolated model components.
Columns (6)--(8): major axes measured in arcmin.
Column (11): position angle in degrees east from north.
Column (16): fraction of total light from bulge component.
Columns (17) and (18): surface brightness and scale length (in arcsec)
of exponential disk component. See text for further details.
}

\end{deluxetable}
 
\clearpage
\endgroup

\begingroup
\clearpage
\begin{centering}
\begin{deluxetable}{@{}l@{\quad}c@{\quad}c@{\quad}c}
\renewcommand{\arraystretch}{0.82}
\tablecolumns{4}
\footnotesize
\tablecaption{Disk Cut-Off Parameters. \label{tbl-4}}
\tablewidth{0pt}
\tablehead{
{No.} & $f_x$ & $\mu_x$ & $r_x$
\\
(1)&(2)&(3)&(4)
} 
\startdata
\phn1&0.12&24.63&\phn2.3\rlap{$''$}\nl
\phn3&0.10&23.27&\phn2.7\nl
17&0.20&24.08&\phn2.5\nl
19&0.40&22.08&\phn4.9\nl
20&0.16&26.02&\phn4.6\nl
29&0.14&24.75&\phn5.0\nl
33&0.23&24.01&\phn4.3\nl
35&0.18&24.52&12.6\nl
54&0.21&24.85&\phn4.3\nl
55&0.35&24.43&\phn4.5\nl
91&0.22&24.73&\phn5.4\nl
\enddata
\end{deluxetable}
\end{centering}
\clearpage
\endgroup

\addtocounter{figure}{-8}
\begin{figure}
\caption{$R$-band optical images centered at the positions of the HI-selected sources.
Each image is $3'\times3'$, and along the bottom is shown the source's
HI profile where the width of the panel corresponds to a 1000 km s$^{-1}$ range.
\label{images}
}\end{figure}

\end{document}